\documentclass{acmart}

\AtBeginDocument{%
  }

\setcopyright{acmlicensed}
\copyrightyear{2025}
\acmYear{2025}
\acmDOI{XXXXXXX.XXXXXXX}
\acmISBN{978-1-4503-XXXX-X/2018/06}

\usepackage{natbib}
\usepackage{placeins}
\usepackage{attachfile}
\usepackage{float}
\usepackage{booktabs}

\usepackage{cleveref}
\crefname{figure}{Fig.}{Figs.}
\crefname{table}{Table.}{Tables.}
\crefname{appendix}{Appendix}{Appendixes}

\begin{document}

\title{Understanding Down Syndrome Stereotypes in LLM-Based Personas}


\author{Chantelle Wu}
\affiliation{%
  \institution{Northeastern University}
  \city{Vancouver}
  \country{Canada}}
\email{wu.chant@northeastern.edu}

\author{Peinan Wang}
\affiliation{%
  \institution{Northeastern University}
  \city{Vancouver}
  \country{Canada}}
\email{wang.pein@northeastern.edu}

\author{Nafi Nibras}
\affiliation{%
  \institution{Northeastern University}
  \city{Vancouver}
  \country{Canada}}
\email{nibras.n@northeastern.edu}

\author{Meida Li}
\affiliation{%
  \institution{Northeastern University}
  \city{Vancouver}
  \country{Canada}}
\email{li.meida@northeastern.edu}

\author{Dajun Yuan}
\affiliation{%
  \institution{Northeastern University}
  \city{Vancouver}
  \country{Canada}}
\email{yuan.da@northeastern.edu}

\author{Zhixiao Wang}
\affiliation{%
  \institution{Northeastern University}
  \city{Vancouver}
  \country{Canada}}
\email{wang.zhixiao@northeastern.edu}

\author{Jiahuan He}
\affiliation{%
  \institution{Northeastern University}
  \city{Vancouver}
  \country{Canada}}
\email{he.jiahuan@northeastern.edu}

\author{Mona Ali}
\affiliation{%
  \institution{Suez Canal University}
  \city{Ismailia}
  \country{Egypt}}
\email{mona@learn2growacademy.com}

\author{Mirjana Prpa}
\affiliation{%
  \institution{Northeastern University}
  \city{Vancouver}
  \country{Canada}}
\email{m.prpa@northeastern.edu}

\renewcommand{\shortauthors}{Wu et al.}
\renewcommand{\shortauthors}{Wu et al.}

\keywords{Persona, Large Language Models, Down Syndrome, Stereotype}


\maketitle

\section{Abstract}

We present a case study of Persona-L, a system that leverages large language models (LLMs) and retrieval-augmented generation (RAG) to model personas of people with Down syndrome. Existing approaches to persona creation can often lead to oversimplified or stereotypical profiles of people with Down Syndrome. To that end, we built stereotype detection capabilities into Persona-L. Through interviews with caregivers and healthcare professionals (N=10), we examine how Down Syndrome stereotypes could manifest in both, content and delivery of LLMs, and interface design. Our findings show the challenges in stereotypes definition, and reveal the potential stereotype emergence from the training data, interface design, and the tone of LLM output. This highlights the need for participatory methods that capture the heterogeneity of lived experiences of people with Down Syndrome.

\section{Author Keywords}
Persona, Large Language Models, Down Syndrome, Stereotype
\section{Introduction}

In the field of human-computer interaction (HCI), personas have long served as a design method to represent users' needs and goals \cite{cooper1999inmates, salminen2022use}, helping designers maintain a human-centered perspective throughout the design and development \cite{blomquist2002personas, salminen2022use}. The recent emergence of LLM-based personas being integrated into design workflows \cite{shin2024understanding} raises the question of how these representations reflect the complexity of users \cite{li2025llm}, particularly when LLMs are used for modeling individuals from underrepresented or marginalized populations \cite{edwards2020three, schulz2012creating}.


Existing approaches to persona development, whether it is data-driven personas \cite{salminen2022use, marsden2019personas} constructed from collected user data or LLM-based personas \cite{sun2025persona, li2025llm}, are prone to the same challenges in addressing the issues regarding truthful representation of needs and goals that is based on lived experiences \cite{salminen2022use, schulz2012creating, edwards2020three}. When personas are based on generalized data or limited stakeholder input, they risk reinforcing oversimplified narratives or stereotypes \cite{schulz2012creating, edwards2020three}. These may include deficit-based language, infantilizing tone, or fixed assumptions about ability, life expectancy, or emotional disposition \cite{schulz2012creating, edwards2020three, enea2012tracking}. Recent advances in LLMs and retrieval-augmented generation (RAG) \cite{lewis2020retrieval} offer new possibilities that enable more nuanced and context-sensitive personas \cite{zerhoudi2024personarag, rizwan2025personabot}. However, these same systems inherit biases from their training data and often produce outputs that reflect dominant cultural scripts, especially when representing disability, race, or gender \cite{bender2021dangers, kotek2023gender, ranjan2024gender}.

In this paper we present a case study of Persona-L \cite{sun2025persona}, a web-based interface that uses LLMs and RAG to create interactive personas of people with Down syndrome. We extend the work by \cite{sun2025persona} to investigate how stereotype mitigation can be integrated into such systems. \cite{sun2025persona} demonstrated how usage of an ability-based framework and LLM in Persona-L can benefit representation of people with Down syndrome in more empathetic and accurate ways \cite{sun2025persona}. However, the extent to which such interface may be introducing or reinforcing stereotypes have not been studied. Moreover, there has been a growing emphasis of the importance of involving experts and those with lived experience as essential partners in the design process \cite{neate2019co, heitmeier2023persona, staley2021should, shin2024understanding}. In particular, regarding Down Syndrome stereotypes caregivers and specialists can apply their knowledge and experience to recognize subtle misrepresentations, challenge inaccurate assumptions, and help ensure that personas reflect the realities of the individuals they are supporting \cite{staley2021should, shin2024understanding}.

To that end our research aim is to answer the following research question (RQ): How can a stereotype detection system be integrated into LLM-based persona generation to mitigate harmful representations, through the perspectives of caregivers and specialists in the Down syndrome community? To address this question, our process involved the following stages: (1) developing a stereotype detection system integrated into Persona-L that flags potentially harmful or inaccurate patterns during persona generation, (2) conducting qualitative validation with caregivers and specialists, including educators and speech–language pathologists, who evaluated generated personas and the detection system.

In answering our RQ, we make three key contributions: (1) Development of a stereotype detection model for persona generation. We designed and integrated a stereotype detection component into Persona-L \cite{sun2025persona} which identifies textual patterns that may reinforce stereotypes. (2) Persona-L evaluation through stakeholder interviews with caregivers and specialists (e.g. special education teachers, speech language pathologists, etc) to assess the quality of generated personas and the accuracy and usefulness of the stereotype detection system, and (3) empirical findings on stereotype emergence, documenting how stereotypes arise both in LLM outputs and in interface design, and highlighting priorities for stereotype mitigation from the perspectives of caregivers and specialists. Our work contributes to the discourse on representational fairness in HCI \cite{Salminen2020RethinkingPersonas}, in particular the use of LLMs for persona creation and the design of systems for underrepresented user groups. With this work we hope to contribute to the human-in-the-loop discourse and chart directions for further development of Persona-L in this direction that would involve caregivers as well as people with Down System in the persona generation framework more closely.


\section{Related Work}

\subsection{Leveraging LLMs to Develop Personas of People with Complex Needs}

Personas in HCI are fictional characters built around user goals, motivations, and behaviours \cite{salminen2022use}, offering a narrative lens that goes beyond “average user” models to ground design in real user needs \cite{cooper1999inmates, chang2008personas, blomkvist2006user}. Personas are especially critical for underrepresented populations with complex needs, where generic personas risk erasing diversity or reinforcing stereotypes unless shaped by lived experiences and intersectional considerations \cite{weber2020using, schulz2012creating, edwards2020three}. Participatory and ethnographic methods address this but are often resource-intensive and difficult to scale \cite{neate2019co, blomquist2002personas, heitmeier2023persona}. To that end, LLMs offer an alternative by simulating human behaviour and enabling scalable persona generation across domains such as science, politics, and healthcare \cite{ge2024scaling, prpa2024challenges, manning2024automated, argyle2023out, horton2023large, wang2024patient}. Recent work shows different ways how LLMs can be applied  to represent people with complex needs: for instance, Persona-L \cite{sun2025persona} builds interactive personas based on ability-based framework. However, \cite{lazik2025impostor} show that while LLM-generated personas can seem informative and consistent, they often lean on stereotypes and lack the contextual and emotional nuance of personas created by humans. This highlights the importance of focusing on diversity and lived experience in persona creation to ensure authenticity and inclusivity.

\subsection{Down Syndrome Stereotypes}

According to Hilton and Von Hippel, “a stereotype is a cognitive structure that contains the perceiver’s knowledge, beliefs, and expectations about a human group” \cite{hilton1996stereotypes}. These cognitive shortcuts enable efficient information processing but often operate automatically and unconsciously, leading to overgeneralized or inaccurate assumptions \cite{hilton1996stereotypes}. Stereotypes are both descriptive (assumptions about how members of a group are) and prescriptive (how they ought to be), and this dual nature plays a critical role in shaping expectations in design and interaction contexts \cite{heilman2012gender}.

In disability contexts, stereotypes are especially significant when they reduce individuals to simplistic narratives such as helplessness or dependency, which ignores the diversity of capabilities and experiences within disabled communities \cite{enea2012tracking}.  Research into the social perception of individuals with Down syndrome reveals a set of consistent and widely held stereotypes, particularly regarding personality traits \cite{enea2012tracking}. These individuals, especially children, are frequently described using predominantly positive characteristics such as being "friendly," "affectionate," "cheerful," and "loving" \cite{carr1995down, collacott1997five}. While these traits are often viewed favorably, they contribute to a "positive stereotype" that, despite its seemingly kind nature, can still be reductive \cite{enea2012tracking}. These characterizations tend to overshadow the individuality and complexity of people with Down syndrome by attributing a narrow emotional range and fixed personality template \cite{enea2012tracking}. Moreover, these positive views coexist with contradictory or even negative societal attitudes toward their inclusion in normative settings such as mainstream education \cite{collacott1997five, wehman2006life}. For example, studies have shown that while many perceive children with Down syndrome as pleasant and cooperative, there remains hesitation about integrating them into regular classrooms, often due to assumptions about their academic limitations or social fit \cite{wehman2006life, copeland1993there}.

\subsection{Stereotypes in LLM-based Personas}

Stereotypes in social contexts are both individual mental models and structurally embedded biases that surface in datasets, interfaces, and algorithms \cite{fiske2020social}. In LLMs, these biases arise from training data \cite{bender2021dangers, blodgett2020language}, algorithmic design \cite{hovy2021five}, and institutional choices \cite{kleinberg2016inherent}, making models especially vulnerable to replicating societal inequalities \cite{jeoung2025examining, shrawgi2024uncovering}. As a result, LLMs often reproduce and amplify stereotypes across gender \cite{lu2020gender, duan2025gender}, race \cite{an2024measuring}, socioeconomic status \cite{liu2024generation}, and other identity dimensions \cite{liu2022quantifying, hofmann2024ai}, posing risks for downstream applications.
In persona generation, these biases can reinforce misrepresentations and narrow assumptions about identity \cite{li2025llm, huang2025survey}, particularly harmful in sensitive contexts such as healthcare or policy design \cite{bender2021dangers, sheng2019woman}. To ensure inclusivity and representational accuracy, recent work stresses the importance of human feedback and co-design practices \cite{shin2024understanding}.

\subsection{Including Caregivers in the Loop for Down Syndrome Stereotype Detection}

Stereotype detection in LLMs has progressed from static metrics such as embedding association tests \cite{shrawgi2024uncovering} to interactive and simulation-based evaluations that reveal situational biases \cite{bai2024fairmonitor}, yet models still fail in domain-specific reasoning and risk defaulting to stereotypes when handling underrepresented communities \cite{li2025large}. To address these limitations, human-in-the-loop approaches \cite{drori2024human, shin2024understanding} are increasingly recognized as essential for scalable and sustainable stereotype mitigation, embedding expert feedback to ensure both technical robustness and social accountability.

Incorporating caregivers into the model development process is both an ethical responsibility and a practical necessity. As long-term participants in the lives of individuals with Down syndrome, caregivers offer context-rich, experiential insights that systems and external annotators typically lack. Participatory research frameworks emphasize that ethically sound and socially relevant systems are best developed when those with lived experience are directly involved in shaping the interventions that affect them \cite{staley2021should}. This is especially important in domains involving marginalized or vulnerable populations, where context cannot be abstracted from experience \cite{staley2021should}.

Caregiver involvement is particularly valuable in tasks such as persona development or user modeling, where large language models (LLMs) are required to represent complex, multidimensional identities \cite{schulz2012creating, edwards2020three, staley2021should}. Recent work comparing human-AI workflows for persona generation found that the most effective method involves humans first structuring and interpreting user data, followed by LLM-generated summaries of that structure \cite{shin2024understanding, smrke2025exploring}. This "LLM-summarizing" approach consistently led to personas that were more representative, empathetic, and aligned with real-world characteristics than those generated without human input \cite{shin2024understanding}.

Caregiver involvement is particularly important when individuals with Down syndrome are unable to provide direct feedback. In such cases, caregiver perspectives serve as a reliable and nuanced proxy. Research using standardized quality-of-life instruments such as the Pediatric Quality of Life Inventory (PedsQL) and IWQOL-Kids has shown that caregivers can accurately report across multiple psychosocial domains, including emotional functioning, physical health, and social engagement \cite{varni2007parent, xanthopoulos2017caregiver}. This suggests caregivers are well-equipped to provide structured, domain-specific input that can inform both model evaluation and refinement.

This contextual sensitivity is especially relevant for stereotype detection, where LLMs often fail to capture the nuanced lived experiences of individuals with intellectual disabilities. For instance, caregivers have been shown to recognize how issues like body image vary in salience between typically developing youth and those with Down syndrome \cite{xanthopoulos2017caregiver}. Their ability to identify such distinctions makes them invaluable in evaluating harms that may otherwise be overlooked by automated systems or annotators unfamiliar with the community.

Including caregivers in human-in-the-loop (HITL) frameworks supports the detection and mitigation of bias, enriches the emotional and contextual grounding of model outputs, and ensures that systems better reflect the complexity, dignity, and diversity of the people they are intended to represent. In this way, caregiver participation contributes to the development of more inclusive, human-centered AI \cite{staley2021should, shin2024understanding, xanthopoulos2017caregiver, varni2007parent, enea2012tracking}.

\subsection{Positioning of Our Research in the Context of Related Work}

Our work builds on Persona-L \cite{sun2025persona}, which introduced an ability-based framework for generating personas of individuals with Down syndrome using LLMs. While the project demonstrated the potential of LLMs to create empathetic and representative personas, it also surfaced challenges related to tone, balance, and the risk of unintentionally reinforcing stereotypes \cite{sun2025persona}. We extend this work by adding stereotype detection module and involving caregivers of individuals with Down syndrome to participate in the feedback loop to evaluate Persona-L output with a goal of refining the system to more accurately reflect perspectives and values of people with Down Syndrome \cite{bai2022training, swamy2024minimaximalist}.
Caregivers offer lived, relational insight that can  contribute to identifying subtle biases and improving the accuracy and nuance of generated content, especially in domains where direct self-reporting may be limited \cite{xanthopoulos2017caregiver}. Our research also draws from recent work on human–AI workflows, which highlights how structuring model input with human expertise can lead to more grounded and inclusive personas \cite{shin2024understanding}. By involving caregivers in this process, we not only aim to reduce harmful or reductive stereotypes in LLM-generated personas, but also ensure these personas reflect the realities of the individuals they are meant to represent.
\section{System Design}

Persona-L was built on these core principles: ability-based design, retrieval-augmented generation (RAG) of LLM output, and balanced representation \cite{sun2025persona}. Grounded in an ability-based framework, the system shifts focus from disabilities to highlighting users' abilities, aligning with self-identities and reducing stereotypes. To ensure contextual accuracy and minimize hallucinations, Persona-L utilizes RAG, which grounds LLM responses in curated, real-world data collected from publicly available sources. Finally, it emphasizes providing a balanced presentation of abilities with constraints and ensuring data transparency and validation. 

\begin{figure}[htbp]
    \centering
    \includegraphics[width=1\textwidth]{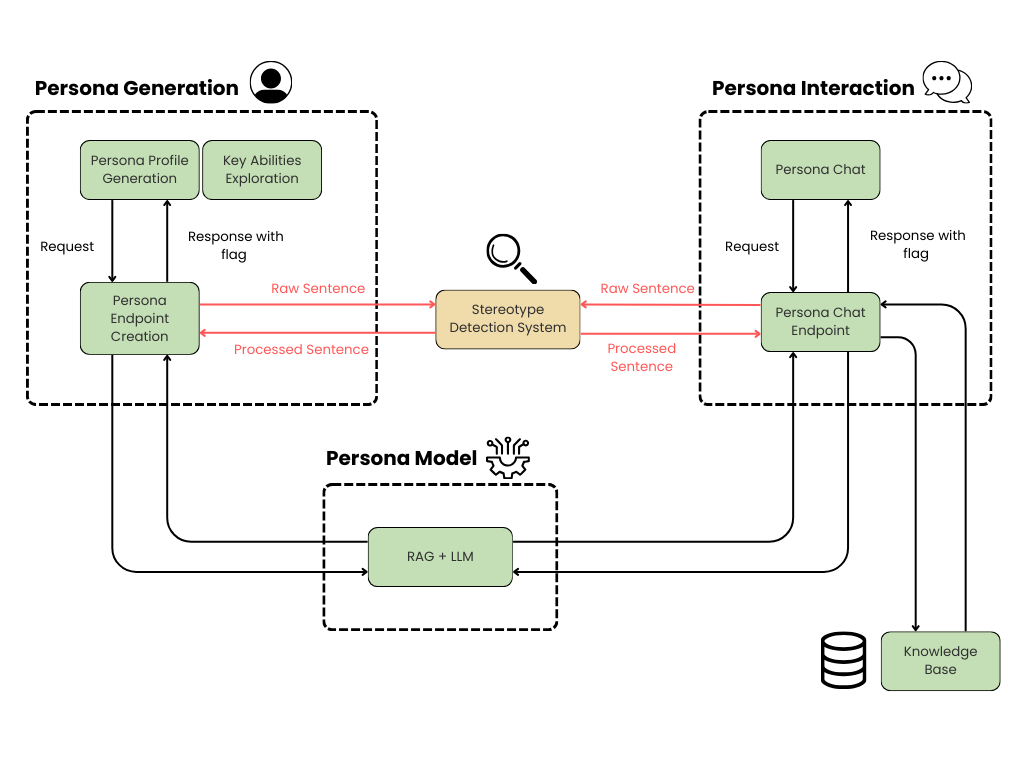}
    \caption{Overview of Persona-L Architecture: components in green were built in the initial Persona-L by \cite{sun2025persona}, spanning both the front-end and back-end architecture; Stereotype Detection module is novel contribution.}
    \label{fig:personal-architecture}
\end{figure}

\begin{figure}[htbp]
    \centering
    \includegraphics[width=0.8\textwidth]{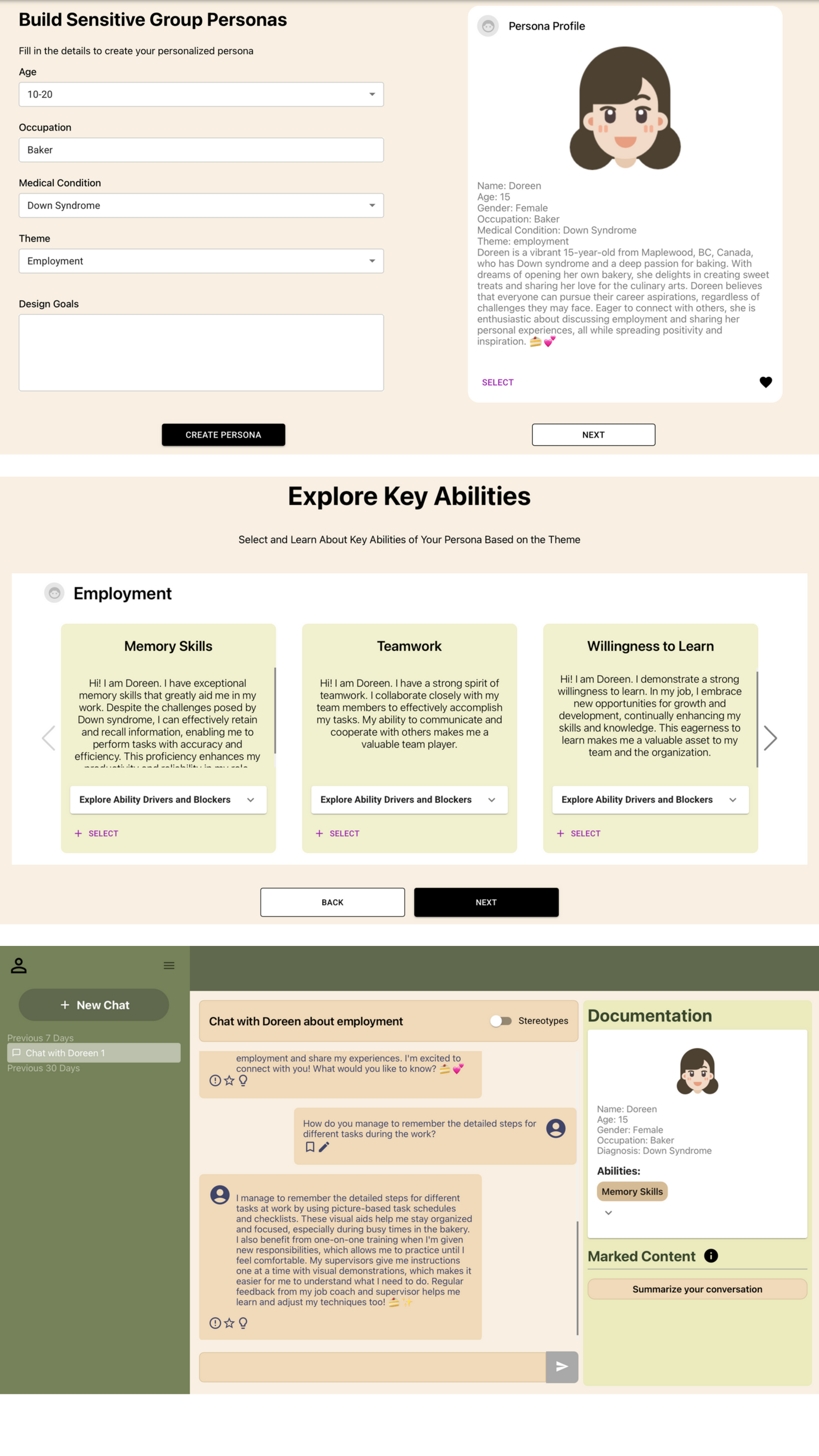}
    \caption{Persona-L UI: Top - Persona Generation; Middle - Key Abilities page; Bottom - Persona Interaction}
    \label{fig:ui}
    
    \phantomsection\label{fig:ui-top}
    \phantomsection\label{fig:ui-middle}
    \phantomsection\label{fig:ui-bottom}
\end{figure}

Initial Persona-L System Design entailed three core modules: Persona Generation, Persona Interaction, and Persona Model (\cref{fig:personal-architecture}). Persona Generation module enables users to create personas by selecting demographic attributes such as age, gender, occupation, and specific conditions, and a theme such as employment, education, or family (\cref{fig:ui-top}--Top Left). According to \cite{sun2025persona}, this design incorporates structured input fields as well as contextual help to guide users in defining personas that capture both challenges and strengths.

After defining initial attributes, users are directed to the Key Abilities page (an ability-based framework)(\cref{fig:ui-top}--Middle), which are sourced from narratives reflecting lived experiences, ensuring authenticity and diversity in persona profiles. Here, users are presented with an option to select specific ability drivers, barriers, and supports relevant to the selected theme. \cite{sun2025persona} emphasized that this phase was built to encourage reflection on a persona’s functional profile, drawing on evidence‑based resources and stakeholder input, and was designed as a stand‑alone module so that future iterations can add new ability datasets without requiring back‑end modifications.

Once a persona and its associated key abilities are finalized, the system transitions into the Interaction Module, where users engage with the persona through the Chat UI (\cref{fig:personal-architecture} — upper right; \cref{fig:ui-top} — bottom). For each message, the conversation with the persona combines the user’s new input with persona metadata, conversation history, and a pre‑populated knowledge base. This retrieval‑augmented setup enables responses that are contextually consistent and aligned with the persona’s defined profile, minimizing generic outputs and supporting scenario‑specific dialogue.

\subsection{Stereotype Detection System}

\begin{figure}[htbp]
    \centering
    \includegraphics[width=1\textwidth]{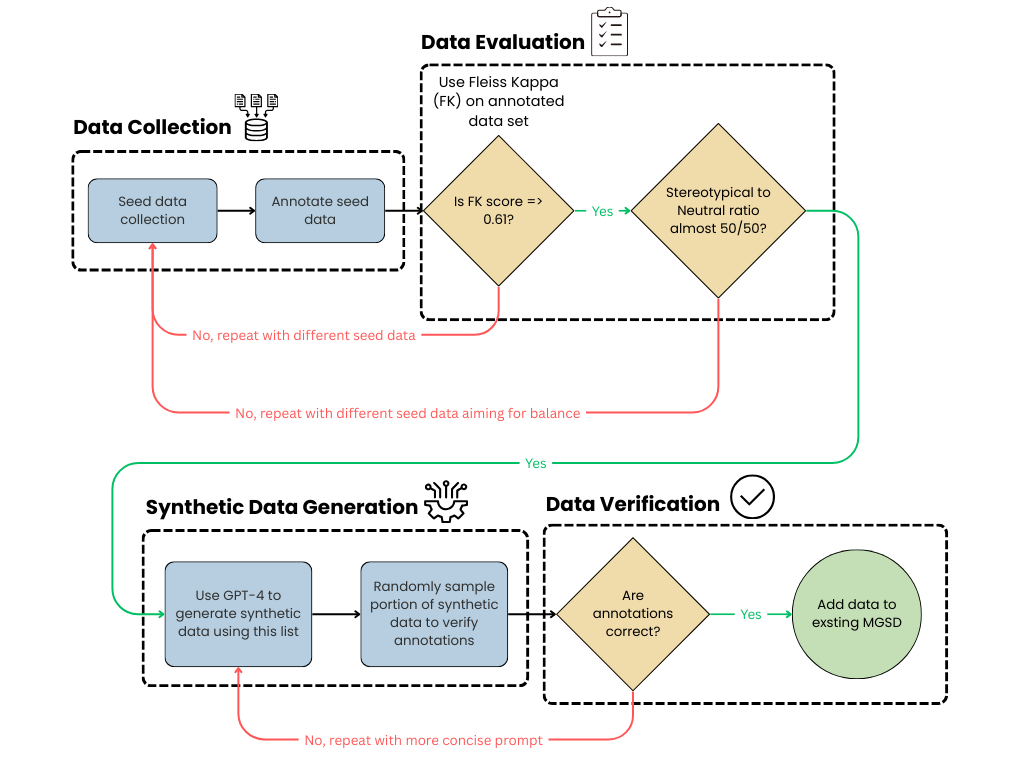}
    \caption{Overview of Stereotype Detection Architecture}
    \label{fig:stereotype-detection-architecture}
\end{figure}

In this study, initial Persona-L architecture was extended to include S\textbf{tereotype Detection Module} (\cref{fig:personal-architecture} — middle). The overall workflow and architectural design of the stereotype detection system are illustrated in (\cref{fig:stereotype-detection-architecture}), which shows the pipeline from sentence input through model ensemble classification to explanation generation.

\subsubsection{Dataset for Stereotype Detection in LLM Output}

To detect stereotypes in Persona‑L’s generated content, a tailored dataset was prepared using the Multi‑Grain Stereotype Dataset (MGSD) \cite{nadeem2021stereoset} as a foundation. The MGSD was originally constructed from StereoSet and CrowS‑Pairs datasets, encompassing over 51,000 labeled examples across categories such as gender, race, profession, and religion. For the purposes of Persona‑L, the scope of the dataset was narrowed to focus on three primary categories relevant to common biases in persona generation: gender stereotype, professional stereotype, and neutral (the baseline category, it captures language that is inclusive, stereotype-free, and doesn't frame identity in a biased way). To address the research goal of capturing stereotypes concerning underrepresented groups, the dataset was extended with examples reflecting stereotypes about individuals with Down syndrome. A seed set of 30 sentences spanning the themes of family, education, and employment was created and manually labeled by three independent annotators, achieving a Fleiss’ Kappa score of 0.775, indicating substantial inter‑annotator agreement \cite{fleiss1971}. Using GPT‑4, the seed sentences were expanded to generate 500 additional instances with careful prompt engineering to maintain diversity and context relevance. The final curated dataset included 16,139 training instances and 1,769 testing instances distributed across four categories: 896 neutral, 647 profession stereotypes, 218 gender stereotypes, and 34 Down syndrome stereotypes. This dataset ensured broad coverage while aligning with Persona‑L’s focus on realistic, contextually nuanced outputs.

\subsubsection{Model Development}

Four transformer‑based language models were fine‑tuned using the extended dataset: ELECTRA \cite{clark2020electra}, RoBERTa \cite{liu2019roberta}, DeBERTa \cite{he2021deberta}, and DistilBERT \cite{sanh2019distilbert}. ELECTRA uses a discriminative pre‑training approach to efficiently learn token replacement detection, RoBERTa builds on BERT with dynamic masking and larger corpora to enhance contextual understanding, and DeBERTa introduces disentangled attention mechanisms that improve semantic representation. DistilBERT, a distilled version of BERT, offers efficiency at the cost of some accuracy. Each model was trained using the AdamW optimizer with cross‑entropy loss, a batch size of 16, and three epochs, with early stopping applied to prevent overfitting. The dataset was tokenized using the respective pretrained tokenizers to maintain architectural compatibility. After evaluation, DistilBERT’s lower accuracy (68\%) led to its exclusion from the final ensemble. The remaining three models were combined using a majority voting mechanism enhanced by probability‑weighted adjustments to form a robust ensemble. The ensemble model achieved an accuracy of 82.06\% and an F1‑score of 82.16\%, with particularly strong results in detecting Down syndrome stereotypes (F1‑score of 99\%), and solid performance across gender and profession categories, although the recall for gender stereotypes remained moderate.

\subsubsection{Integration of the Stereotype Detection Model into Persona-L System}

The trained ensemble model was integrated into Persona‑L’s back‑end workflow as a post‑processing layer for all LLM outputs. After the create‑persona or persona‑chat endpoint generates a narrative or a response, the output is segmented into individual sentences and passed through the ensemble detector. Each sentence receives a classification label: gender stereotype, profession stereotype, or down syndrome stereotype, along with a confidence score. These results are bundled with a GPT‑4‑generated explanation and returned to the front‑end. In the interface, flagged sentences are highlighted and augmented with tooltips that display the confidence score and explanation text, allowing users to quickly assess and interpret potential bias without automatic correction. This integration ensures that the stereotype detection feature supports users in evaluating content while preserving their control over persona development.

\begin{figure}[htbp]
    \centering
    \includegraphics[width=1\textwidth]{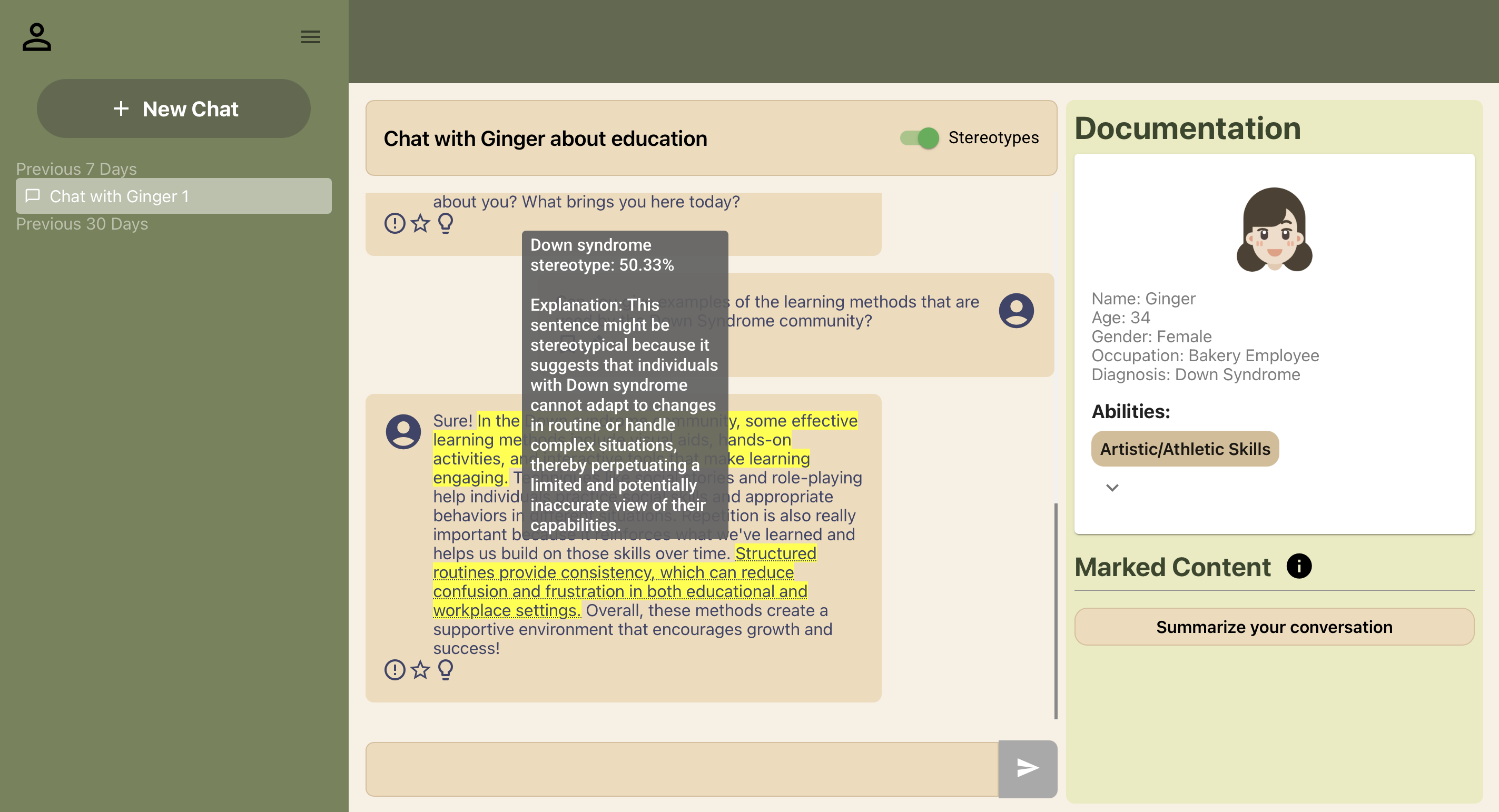}
    \caption{Example of Stereotype Flagging in Persona-L}
    \label{fig:stereotype-detection-ui}
\end{figure}

In addition to classification, GPT‑4 was used to generate concise natural‑language explanations for any flagged stereotypes, providing users with transparent reasoning (\cref{fig:stereotype-detection-ui}). These explanations clarify why a sentence was identified as stereotypical, referencing specific bias patterns. Flagged sentences are visually highlighted within the Persona-L interface, with tool tips providing users with both the confidence score and the generated reasoning. This explanation allows users to evaluate and make decisions based on the flagged content. The system is intentionally developed to be non-intrusive, offering detection without automatic correction. This leaves the decision-making process to the end user \cite{cheng2023marked, gilardi2023chatgpt}.

\section{User Study}

\subsection{Participants}
This study involves family and professional caregivers, speech–language pathologists, and behavioral therapists with direct experience supporting individuals with Down syndrome. Participants, who are 18 years or older, fluent in English, and have at least one year of experience providing care were purposively selected to ensure contextual familiarity with the personas being represented in Persona-L, aligning with established qualitative research practices \cite{plano2017mixed}. Participants were recruited through personal contacts, snowball sampling, and online outreach on LinkedIn and Reddit. A total of ten participants (N=10) took part in the study. Participants’ demographic details are provided in (\cref{Table:participant-table}). Prior to the study, all participants reviewed the study description and signed informed consent forms. The study was approved by the Institutional Review Board (IRB).

\subsection{Study Procedure}

This study followed a sequential, convergent mixed–methods design \cite{creswell2021concise}. Each session lasted approximately 60–80 minutes and proceeded in the following order.

\subsubsection{Introductory Questions}
The session began with introductory questions about participants’ professional or caregiving background and their experiences with individuals with Down syndrome. Participants described the nature and duration of their role, the settings in which they worked (e.g., home, clinic, school, or community), and the types of communication needs and preferences they most frequently encountered. They also reflected on how communication styles vary across family, educational, and employment contexts, and shared insights or misconceptions they wished more people understood.
Participants were additionally asked to define what they considered a stereotype, share common stereotypes they had observed, and explain how they addressed such stereotypes in practice.

\subsubsection{System Introduction and Persona Interaction}
Participants were introduced to Persona-L, a web-based prototype for generating AI-driven personas of individuals with Down syndrome \cite{sun2025persona}. Each participant created three personas, each grounded in a distinct communication context (family, education, or employment). After creation, they selected one persona to explore in depth, engaging with it through the Persona Generation, Key Abilities, and Chat UI interfaces (\cref{fig:ui-top}).
Participants asked the persona questions about communication preferences, routines, day-to-day experiences, and support needs.

\subsubsection{Quantitative Survey}
After interacting with the persona, participants rated their agreement with six statements on a 5-point Likert scale (1 = Strongly Disagree, 5 = Strongly Agree) \cite{varni2007parent, xanthopoulos2017caregiver}. The items assessed persona accuracy, realism, balance of strengths and challenges, and trustworthiness for design use. Questions for this survey can be found in (\cref{Table:likert-questions.png}).

\subsubsection{Semi-Structured Interview}

A follow-up interview reviewed participants’ survey responses in greater depth. Participants elaborated on which aspects of the persona felt accurate or relatable, how realistic the communication style seemed, and whether any important qualities or behaviors were missing. They also reflected on mismatches in tone, highlighted moments that felt authentic or stereotypical, and suggested refinements to better represent individuals with Down syndrome. Participants also commented on design choices in the interface, providing feedback on how interaction flows could better support reflection and avoid confusion.

\subsubsection{Stereotype Validation}
The evaluation concluded with a stereotype validation task.
Independent review: Participants revisited persona responses and identified any stereotypical or problematic statements on their own.
System-assisted review: Persona-L’s built-in stereotype detection system was enabled, and participants reviewed the flagged content. They judged whether flagged items were indeed stereotypical, discussed why they agreed or disagreed, and suggested alternative phrasings.
This process allowed us to compare human and system definitions of stereotype, surfacing both subtle cases of bias missed by the model and instances where the model was overly sensitive. Caregivers provided nuanced guidance on how stereotypes intersect with daily contexts like employment or education, and highlighted how seemingly benign sentences (e.g., infantilizing tones or exaggerated cheerfulness) could still perpetuate reductive framings. They also identified design needs in the interface, such as clearer differentiation between neutral and potentially stereotypical statements.

\section{Findings}

\subsection{Caregivers’ perspectives on Down syndrome stereotypes}

When asked to define a stereotype, participants consistently described it as an oversimplified generalization imposed on individuals based on group characteristics rather than personal attributes. P1 explained that a stereotype is “something that people put onto other people of a similar group … by some sort of characteristic rather than by actually getting to know them.” Similarly, P10 referred to them as “a generalised judgement based on a very relatively small number of things,” while P4 defined them as “broad assumptions of a person based on their diagnosis, look, and behaviour.”

Additionally, caregivers emphasized that stereotypes often persist not only because of unfamiliarity but also because of cultural and social reinforcement. P9 highlighted how such assumptions are frequently passed down: “They assume common knowledge that’s believed to be true just by what you hear ... from your parents or friends or peers.” P7 added that stereotypes can arise both from direct experience and from ignorance: “Assumptions or stereotypes come from growing up in a certain place, and also from being uneducated about certain groups like people with disabilities.”

Moreover, several caregivers described stereotypes as visually driven, particularly in the context of disability. P5 noted that stereotypes can be shaped around assumptions of “how a person acts, appears, or looks.” P8 illustrated this point through her own experience managing her child’s appearance: “When he went to high school, we got him a Letterman jacket ... If they look more appropriate and similar to their peers, there’s a better chance they’re going to be accepted.” For her, conforming to appearance was a way to reduce exclusion for her child.

Some participants also reflected on the presence of “positive” stereotypes about Down syndrome. P2 explained that these assumptions, while seemingly harmless, still diminish individuality: “All they know is somebody’s happy and they get along and they’re happy to hug or something superficial like that.” P1 also resonated with this point, acknowledging that both positive and negative stereotypes but stressing their reductive nature: "t’s usually negative. Well they can be positive too ... but it’s still saying somebody has a certain trait because they belong to a certain group.”

Overall, participants framed stereotypes as cognitive shortcuts; socially reinforced, frequently visual, rarely accurate. As P6 summarized: “Stereotypes are often simplified and generalised ideas ... definitely not applicable to individuals within that group.” Altogether, the caregivers' reflections revealed a wide range of how stereotypes were defined and understood. This diversity of perspectives highlights how difficult it is to arrive at a single consensus definition of what a stereotype means to even the people closest to people with Down syndrome. More in-depth results can be found in  Appendix (\cref{sec:most-common-ds-stereotypes})

\subsection{Complexity of Down syndrome and the challenge to represent complexity in data}

Across interviews, participants highlighted the wide variability in cognitive ability, communication, behaviour, health conditions, and learning styles. This variability questions the feasibility of fixed or static personas commonly used in design practice. As per P5, "there is no one-size-fits-all." P2 further explains this heterogeneity: "People with Down syndrome have intellectual disabilities. They have physical disabilities. Some of them can't communicate at all." Her point was not intentionally reductive, but rather emphasized that variability is inherent to the condition. As well from a clinical perspective, P2 noted that, "Trisomy 21 affects a lot of different systems. It's a single gene, which means a ton of things." She compared this with diagnoses like cerebral palsy, which may not be present as a localized physical trait. Comparatively, Down syndrome affects cognition, motor control, speech, social interaction, and overall health; variation is the norm rather than the exception. In terms of diversity of communication, P6 stressed that, "Some people with Down syndrome maybe really just talk in one phrase... it's very hard to pinpoint, 'this is how they would be.'" In her school-based practice, she had encountered students who could only speak in one or two words, while others were far more verbal and independent. 

These types of differences ultimately reinforce the difficulty of creating representative personas as a single or even multiple personas may unintentionally oversimplify meaningful distinctions. P5 illustrated this with an example: her son and his friend developed an application for children with Down syndrome, but their trials showed that "there was no one size fits all." The technology worked for some but not others due to fundamental differences in cognition and physicality. For Persona-L, participants recognized the attempt to address this complexity through customization of personas by turning to the ability-based framework. However, participants still felt the current system lacked depth. Although some participants expressed that some dialogue elements felt relatable, other participants felt that the personas were stereotypical and oversimplified. This included, "Not giving the real-life feel of how an actual Down syndrome person is" (P7) as the persona's responses felt repetitive, cheerful, and not representative of the challenges people with Down syndrome face. Moreover, P2 added that the personas failed to specifiy which kinds of support needs or Down syndrome profile were being represented: "By saying that they have Down syndrome, I don't know what that means... Were we talking about somebody with physical disabilities? Were we talking about somebody with speech issues? I don't know that without further clarification." Overall, participants felt that although there was an attempt to represent the complexity of lived experiences in Persona-L, Down syndrome is extremely complex and requires further consideration for persona creation.

\subsubsection{Challenges at the interface: defining Down syndrome persona characteristics}

The participants’ first task was to create personas using Persona-L that reflect the demographics of people with Down syndrome they took care of. This was also an opportunity to collect participants’ feedback regarding how demographic parameters such as age, occupation, and personal identity descriptors, could potentially be reinforcing limiting views of individuals with Down syndrome. 

\textit{Age}: In Persona-L the highest age was set to 40 (because the data collected covered experiences of people with DS between 10 and 40 years old). However the participants understood this as a failure to reflect the current life expectancy and reality of individuals with Down syndrome. Several participants highlighted the implicit bias in capping age in the 30s or 40s, suggesting this reinforces outdated medical assumptions about life expectancy. P1 expressed surprise and concern at the interface’s limitations: “People with Down syndrome can live into their forties, sometimes even fifties.” P8 echoed this sentiment, emphasizing medical advancements and shifting norms: “They're living a lot longer nowadays, so that's kind of interesting.” Such constrained demographic options were interpreted not simply as data limitations, but as manifestations of societal assumptions about shortened lifespans, effectively enforcing the stereotype that people with Down syndrome are not expected to live full, long lives.

\textit{Occupation}:  Participants described the occupational categories as either too aspirational and unrealistic or unimaginatively restrictive, revealing the tension between aspirational inclusion and realistic representation. Some participants flagged job titles like “Customer Service Professional” or “Waiter” as potentially misaligned with common abilities by individuals with Down syndrome. As P1 stated, “You have to be very high functioning Downs to do that.” She suggested more appropriate alternatives such as “Customer service persona” or “Part-time community centre persona,” emphasizing that supportive roles, rather than independent high-pressure ones, might better reflect reality for many individuals. P8 provided further nuance by highlighting that many individuals transition into day programs after school: “A lot of our kids, once they age out of school at 22, they go into what they call day programs ... A good majority of them will not work in an employment type of situation.” This feedback illustrates that occupation choices within interfaces can act as mirrors of broader social attitudes, either reinforcing stereotypical limitations or imagining new but unrealistic possibilities. 

\subsubsection{Highlighting abilities must be balanced with challenges}

The Key Abilities in Persona-L was designed to reframe Down syndrome around strengths, preferences, and individuality rather than deficits. While participants appreciated this intention, they also raised concerns that ability-based framework may still introduce limiting or idealized portrayals. Participants emphasized that abilities are not uniform across individuals with Down syndrome; context, environment, and support structures play critical roles. P10 explained that qualities like persistence may be internally driven for some children but externally supported for others. As well, caregivers noted that certain traits while aspirational simply did not resonate well with individuals with Down syndrome, such as teamwork. P4 noted, “Teamwork is never strong in my experience with any kids that I've worked with [with] Down Syndrome.” She also doubted whether advocacy applied broadly: “I never think of my Down Syndrome clients or brother as good advocates for themselves.” Some other traits such as "willingness to learn" or "memory skills" were also seen as more context dependent. P9 pointed out: “They might be willing to learn, but motivation and capabilities might depend on the person.” Similarly, P4 felt that while memory skills resonated, teamwork and willingness to learn did not always align with her experience.

As well, many caregivers felt the personas presented an overly positive image, overlooking everyday challenges, emotional fluctuations, and systemic barriers. P1 reminded that “Every human being ... we all have things we don’t like. We all have bad days and bad moods,” pointing out that omitting these realities flattened the persona’s emotional depth. P2 echoed this critique: “I don’t think it really talked about any weaknesses ... it didn’t talk about struggles.” Participants responded positively to attributes such as “persistent,” “curious,” or “creative,” describing them as resonant and empowering. P10, for instance, agreed: “The qualities you guys put out, persistent and creative... that stuff does resonate with me.” At the same time, she stressed the need for moderation: “It’s kind of finding that happy medium, right?... It’s just figuring out that sweet point.” For the need of emotional nuance, participants suggested certain characteristic customizations that could be incorporated in the system to further improve it. P10 suggested traits capturing social energy or emotional regulation, such as “conversations energize me” versus “conversations deplete me.” She further recommended considering questions like: “What brings me comfort or what is calming? How do I get frustrated really easily?”

\subsection{Evaluation of Stereotypes Emerging from the Interaction}

\subsubsection{RAG improves content accuracy}

Participants widely agreed that the persona’s responses often reflected accurate information about Down syndrome, particularly in relation to interests, strengths, and daily routines. However, they also identified a consistent gap between factual correctness and contextual authenticity. While the persona conveyed accurate content, participants wished to see more personal, detailed accounts of lived experiences. Several participants acknowledged that the persona captured common cultural or behavioural reference points associated with individuals with Down syndrome. For example, P10 noted that “the Taylor Swift thing was pretty accurate,” explaining that pop culture references like movies or sports tend to be “big experiences” for many of the students they work with.

Additionally, participants acknowledged that the persona provided accurate and realistic representations of the contextual supports commonly used by individuals with Down syndrome. These included references to educational aides, visual schedules, checklists, and social skills development strategies; support mechanisms that participants recognized as helpful and appropriate. P6 described the persona’s portrayal of support systems as well-aligned with their expectations and experiences. “I do think that the strategies that you guys put in, I would expect that someone with Down Syndrome would benefit from,” they explained, emphasizing that resources such as “a visual schedule,” “an educational persona,” and “checklists” were all practical accommodations regularly used in educational and therapeutic settings. They also noted that more socially oriented tools like “a social skills group” were fitting additions, describing them as both “beneficial” and something they “would expect them to potentially benefit from.” P1 also pointed to moments where the persona provided contextually accurate advice, even if the speaker in the scenario did not seem likely to express it that way themselves. “The down person would never say this,” they said of one particular recommendation, “but this is actually the right advice.” This response suggests a partial success: the persona understood what kinds of supports were needed, but still struggled to match them to a believable speaker voice or delivery style. As a result, contextual alignment did not always translate into perceived authenticity.

Other participants highlighted that the content, while rooted in correct generalizations, lacked narrative depth or individual specificity. P2 suggested that “some things were accurate, but they were very narrow in how they were accurate.” They elaborated that the persona often repeated the same shallow points, giving the impression that “there wasn’t a lot of data there,” and that the representation was “just kind of one note.” Although the persona appeared to pull its responses from familiar sources, this contributed to a sense of artificiality. As P2 put it, “it definitely looked like it scraped websites that talk about Down syndrome,” referencing advice such as “use visual aids, use cues,” which they noted felt like what was found in learning resources, not insights derived from actual lived experience.

Another subtlety identified by participants was the implication that of exaggerated positivity and a limited range of emotions. While the persona was intended to be warm and optimistic, some participants felt that this came at the cost of emotional realism; flattening the character’s affect and inadvertently reinforcing reductive assumptions about emotional capacity. P1 pointed out that “everything is just so pleasant and nobody would say this.” She argued that this uniform positivity contributed to a stereotype that individuals with Down syndrome lack emotional complexity, noting that the persona’s speech “simplifies the range of emotions that the person with Down Syndrome has.” In her view, this not only misrepresented the lived experience of people with Down syndrome but also denied them the full spectrum of emotional expression. 

Participants also noted the overly polished language, but despite this, the persona still successfully captured elements of the lived experience of people with Down syndrome. P1 remarked that certain behavior portrayals were true even when the speech patterns didn’t quite match. Recalling a moment when the persona encouraged participation in cooking, P1 commented, “The content I think is accurate. Cooking? Yeah. My stepbrother likes to help out with cooking. He won’t say he does, he won’t say it, but if you say you want to help out in the kitchen, he’ll jump over and do it.” While the wording in the persona's output might have been unrealistic, the motivation and action behind it (volunteering in the kitchen when prompted) was seen as an accurate reflection of lived experience. This theme appeared elsewhere in their feedback as well, as P1 noted that while certain responses might sound unlikely coming directly from someone with Down syndrome, “this is actually the right advice.” Altogether, participants found the persona’s content to be largely accurate in terms of categories and characteristics, but lacking in depth, individuality, and developmental nuance. Factual correctness was not enough to establish authenticity; instead, participants wanted representations that moved beyond generic affirmations and into more personal, dynamic portrayals. 

\subsubsection{Structure: Persona-L responses can be seen as overly lengthy and scripted}

While participants acknowledged that certain aspects of the persona’s content were grounded in facts, many felt that the responses failed to reflect the realities of how individuals with Down syndrome communicate, process emotions, or express themselves. The mismatch, they emphasized, was not always in what was said, but in how it was structured and delivered.

Across multiple interviews, participants expressed concern that the persona's delivery of speech felt overly polished, complex, and sounding scripted and unrealistic rather than spontaneous. P3 explained: “It was just way too many words ... the sophistication level of this speech was too high.” For them, the persona’s tone was more suited to someone “applying for a job” than to a neurodivergent child in a casual conversation. They elaborated: “It sounded more like interview questions... like, ‘I like this, and the reason I like this is because dah, dah, dah.’” P3 also observed the repetition of an unnatural structure: each response followed a predictable “A, B, C” pattern: an answer, a personal explanation, and a follow-up question. While organized, this formula made the dialogue feel mechanical: “That’s not how speakers [with Down syndrome] speak.”

Other participants similarly stressed that the persona’s responses were excessively lengthy. P1 called for “shorter responses, less complexity,” and noted the lack of natural turn-taking: conversations should feel more reciprocal and interactive. P10 agreed, remarking: “I don’t think they would respond in that length or that detail.” Instead, responses from individuals with Down syndrome are often simpler and more general: expressions like “we hang out” or “we play basketball” rather than long, elaborated narratives. P3 added a physiological perspective: many individuals with Down syndrome lack the breath control or core strength to sustain long sentences. In her view, the persona’s output was “at least three standard deviations higher” than realistic speech expectations, making the responses not just stylistically but physically unrealistic.

In addition, several caregivers highlighted mismatches between the persona’s age and its language style. The persona for some participants was modeled as a 12-year-old, but these participants felt the responses far exceeded both chronological and cognitive expectations. P5 noted this gap: "It’s insensitive to the fact that it’s a 12-year-old boy with Down syndrome... too much fluff.” She emphasized the importance of simpler, accessible responses: “No 10-letter words... you can simplify something and make it easy for people to understand.” She further noted that “some children with Down syndrome… wouldn’t be anywhere close to their chronological age,” pointing to the risks of assuming developmental parity where it does not exist and challenges of translating it to accurate speech patterns.

\subsubsection{Delivery - Oversimplification of persona through happy, positive attitude}

Participants frequently highlighted the persona’s overwhelmingly cheerful and upbeat tone as a core stereotype embedded in its delivery. While individuals with Down syndrome may be joyful or expressive, participants emphasized that portraying them as perpetually happy erases emotional nuance and enforces a reductive, unrealistic view of their lived experiences. Across interviews, this “always happy” tone was described as a form of emotional flattening, where all responses were marked by excessive positivity, regardless of the context. P1 found this tone not only stereotypical, but also infantilizing, explaining that “everything has an exclamation point. Everything is happy.” She noted one specific example, “I think it’s great, and I don’t have a brain!” as a moment where the persona felt “simple in a not intellectually engaged kind of way.” P2 made similar observations, noting that “everything was positive,” even when the content didn’t call for it, and that this constant enthusiasm made the persona’s communication “rosy” in a way that didn’t align with her understanding of Down syndrome. For P10, the portrayal erased important aspects of emotional range, stating that while many individuals “respond well to play interaction,” they are “not always happy,” and may experience frustration, sadness, or disinterest just like anyone else. 
Other participants echoed this, identifying how the persona’s constant enthusiasm risked reinforcing a narrow and misleading cultural narrative. P7 was noted the assumption that “they’re always happy,” warning that such portrayals undermine the complexity of real individuals. “I felt like they just kind of repeated themselves,” she noted, referencing how the persona reiterated phrases about liking music, dancing, and family support without ever expressing any struggle or ambivalence. P9 similarly commented on the emotional tone, saying it “felt very excited” in an unrealistic way. They noted that in practice, individuals might be more reserved or neutral, especially when discussing topics like school or physical education: “Maybe in real life it’s more like, ‘yeah, I like gym, I play soccer.’”

Across participants, there was a consensus that the persona’s consistently upbeat delivery represented a stereotype, one that many felt does more harm than good. As P2 concluded, while being cheerful is not inherently problematic, packaging an entire persona around that single emotion strips away authenticity and suggests that individuals with Down syndrome are emotionally one-dimensional. Participants urged a more balanced tone that allows space that reflects the real range of moods and emotional expression they regularly encounter.
\section{Discussion}

In this study we set out to explore the following research question: How can a stereotype detection system be integrated into LLM-based persona generation to mitigate harmful representations, and how can its effectiveness be validated through the perspectives of caregivers and specialists in the Down syndrome community?

Our findings revealed both the opportunities and limitations of combining RAG with stereotype detection in the context of Down syndrome personas. While the system successfully improved factual grounding and revealed potentially stereotypical outputs, caregiver evaluations revealed that LLM personas can introduce stereotypes in less obvious ways- at the Interface level, and through Interaction with LLM-based persona. We discuss these points in the following sections.


\subsection{Caregivers’ Perspectives on Down Syndrome Complexity and Stereotypes}

Our work on the evaluation of potential stereotypes present in Persona-L output highlighted the necessity for a clear definition of stereotypes related to Down Syndrome. Our findings show that some participants described stereotypes as socially reinforced shortcuts that obscure individuality, which resonates with \cite{hilton1996stereotypes} framing of stereotypes as overgeneralized cognitive structures. While some noted the persistence of “positive” stereotypes such as assumptions of perpetual cheerfulness, caregivers stressed that these were just as limiting as negative ones because they erase emotional nuance and adversity. Prior work similarly shows that these portrayals reduce identity to simplified, familiar images rather than acknowledging diversity \cite{enea2012tracking}.

Caregivers emphasized that Down syndrome is inherently complex due to its wide variability in communication, cognition, health, and behaviour. From their perspective, functional ability does not necessarily map onto diagnosis, as individuals may present very different strengths, challenges, and support needs. HCI literature similarly emphasizes that lived experience is inherently complex, multidimensional, and embodied which requires methods that move beyond surface-level descriptions to include varying dimensions of lived experiences. \cite{prpa2020articulating} highlight that articulating such experiences require methods sensitive to subtle, situated detail, since conventional representations often miss what is “lived and felt” in the moment. This resonates with caregivers’ insistence that individuals with Down syndrome embody diverse and evolving patterns of communication, emotion, and social engagement. 

Caregivers observed that portrayals of Down syndrome in public narratives and in personas often take on a celebratory or one-dimensional tone that obscures struggle and adversity. Prior work has shown that data-driven persona approaches in HCI such as using clickstreams or social media to construct archetypal users offer scalability but often reduce complexity into stereotypical archetypes \cite{zhang2016data, jung2017persona}. Our results found a similar limitation: while using publicly available narratives and sources provided factual accuracy and a foundation for persona creation, the resulting portrayals lacks depth of lived experiences. Due to this, caregivers' feedback indicates the need for more participatory and crowdsourced mechanisms where individuals with Down syndrome and their caregivers can be humans-in-the -loop and directly contribute to the narratives describing experiential complexity of daily life. There are early precedents in this direction though they also necessitate more careful deliberation of best practices. \cite{Shaffer2025, bryen2016ethical} describe the challenges and ethical considerations of directly involving individuals with intellectual disabilities and Down syndrome in research participation. This includes consent capacity, assent procedures, accommodations for communication differences, and safeguarding respect and autonomy. All of these are imperative to ensure that these systems reflect complexity and maintain ethical integrity.

\subsection{Stereotype at the Interface}

While much attention has focused on how LLMs encode and reproduce stereotypes \cite{bender2021dangers, kotek2023gender, shrawgi2024uncovering, an2024measuring}, our study demonstrates that interface design itself can embed and perpetuate stereotypes. 
This resonates with HCI literature showing how design decisions can unintentionally project biases and even potentially manipulate user expectations. \cite{sanchez2023ethical} warn that even when an interface is framed as persuasive or user-friendly, it can blur into manipulation to subtly shape user perception by limiting options, obscuring alternatives, or presenting information in unbalanced ways. 

In Persona-L, caregivers pointed out that capping the age range at 40 reinforced outdated assumptions about shortened lifespans, while occupational categories were either overly aspirational or too narrow, failing to reflect common employment trajectories in adulthood. These constraints illustrate how even “neutral” parameters can encode normative assumptions about disability, reducing representation rather than expanding it. This illustrates what \cite{sanchez2023ethical} described as the tension between persuasion and manipulation. The line is crossed when design choices reduce user autonomy by embedding stereotypes.


In addition, the ability-based framework was flagged by the caregivers for potentially introducing unbalanced views of Down Syndrome. While its intent was to reframe personas through strengths and preferences \cite{sun2025persona}, caregivers observed that the absence of more nuanced traits could compromise the complexity of lived experiences. To that end, caregivers described it as creating an idealized picture that flattened individuality. \cite{sanchez2023ethical} emphasizes this risk as persuasive designs often leverage positive framings and emotional appeals, but when such framings mask difficulties or omit context, they edge toward design that conceals reality. 

Overall, stereotype mitigation in AI personas is not solely a matter of refining model outputs, but also requires critical attention to interface design choices. As our findings suggest, even well-intentioned designs can inadvertently reproduce outdated assumptions or flatten the diversity of lived experiences. This highlights the importance of participatory practices that include caregivers and individuals with Down syndrome as essential in all stages of persona creation.

\subsection{Stereotype Emergence from the Interaction}

While much of the literature on stereotypes in LLMs has focused on biased content \cite{bender2021dangers, kotek2023gender, shrawgi2024uncovering, an2024measuring}, our study demonstrates that stereotypes may also surface through conversation delivery of the persona. Caregivers emphasized that while factual content was accurate through the integration of RAG, the tone, structure, and emotional framing of the responses introduced forms of stereotyping. This finding extends prior work that highlights how bias can operate not only in what AI systems say but in how they say it, including rhythm and formality of generated outputs \cite{sourati2025homogenizing}.

Participants acknowledged that the use of RAG improved factual accuracy. The persona drew appropriately on reference materials, informing the response to include best practices when asked for communication aids such as visual schedules, educational aides, and social skills groups. These were widely recognized by caregivers as practical strategies familiar from daily life. However, caregivers also noted that accuracy in content did not guarantee authenticity in delivery. Participants found the way Persona-L conveyed the information was too elaborate, formal, or rehearsed to resemble how an individual with Down syndrome would likely communicate. Another recurring theme was the perception that Persona-L’s responses were overly scripted, lengthy, and formula-like. 



These observations align with prior concerns that LLM-generated dialogue often lacks spontaneity, producing formula-like speech that flattens natural variability \cite{sun2025persona}. For caregivers, this mismatch went beyond style to include developmental inaccuracy: the persona’s speech was consistently more advanced than what they would expect from a child of the same chronological age with Down syndrome, showing that delivery may implicitly deviate from lived experiences from unrealistic competence.


Caregivers also identified the persona’s consistent positivity as a key stereotype embedded in its delivery. While individuals with Down syndrome are often characterized as friendly or cheerful \cite{enea2012tracking}, participants critiqued the persona’s constant enthusiasm as an oversimplification that erased frustration, sadness, and the spectrum of emotions individuals with Down syndrome have. This “always happy” stereotype documented in the literature \cite{carr1995down, collacott1997five}, was reproduced not through factual inaccuracies but through the framing of dialogue. Similar concerns have been raised in broader HCI research, where tone-polished AI systems risk giving the illusion of empathy while concealing emotional nuance \cite{sourati2025homogenizing}. For participants, this suggested that individuals with Down syndrome are emotionally one-dimensional, a stereotype that directly undermines their dignity and complexity.


In summary, while RAG improved the accuracy of persona content, caregivers highlighted that stereotypes emerged through the delivery of responses. This indicates that addressing these issues not only requires grounding content in accurate sources but also developing tone-appropriate approaches that align with conversational rhythms, developmental profiles, and emotional variability that ensure that personas reflect lived experience rather than producing oversimplified portrays. 
\section{Future Considerations}

\subsection{Persona-L to present views of caregivers/therapists on the needs of people with Down syndrome}
Persona-L shows potential not only as a tool for generating personas of individuals with Down syndrome, but also as a resource for designers, educators, and therapists seeking to understand the perspectives of caregivers. Importantly, when creating resources for design teams, particularly those who may have limited prior experience interacting with individuals with Down syndrome; the choice of voice becomes critical. While fully mimicking the language or communication patterns of a person with Down syndrome could offer authenticity, it may also make the persona less accessible or harder to interpret and translate into design, 
In contrast, representing the perspectives of caregivers or therapists provides a more easily understandable entry point, framing the lived realities of Down syndrome through voices that are both contextually grounded and more immediately legible to design professionals. Future iterations could therefore provide persona views that intentionally layer caregiver or therapist perspectives alongside those of individuals with Down syndrome, allowing design teams to navigate this ecosystem of relationships in ways that balance authenticity with accessibility.
\subsection{Tuning Persona-L to match authenticity of language patterns of people with Down syndrome}
In contrast, there are situations where closely mimicking the language patterns of individuals with Down syndrome becomes especially valuable. While retrieval augmentation improved factual accuracy, caregivers underscored that delivery often lacked the cadence, brevity, and emotional variability characteristic of real communication. Future development should therefore focus on matching the rhythms, styles, and conversational patterns of individuals themselves, ensuring that generated personas reflect not only what is said but how it is said. Such modeling has strong potential as a training or educational tool. For example, new caregivers, therapists, or educators in the preliminary stages of learning how to interact with people with Down syndrome could benefit from practicing with personas that realistically mirror authentic dialogue rhythms. This can help build familiarity, reduce initial discomfort, and prepare them to engage more effectively and empathetically in real-world scenarios. Whereas caregiver and therapist perspectives support accessibility and interpretability for design teams, personas that capture authentic communication styles function as experiential training resources—together representing complementary dimensions of how Persona-L can support understanding and inclusion.

\subsection{Crowdsourcing}
To address the challenge of capturing the inherent complexity of Down syndrome, future work should explore platforms where individuals with Down syndrome, their caregivers, and their support networks can contribute lived experiences directly. Rather than relying solely on publicly available or researcher-curated datasets, a crowdsourcing model would allow for continual enrichment of the persona library with diverse perspectives. This approach could be implemented through structured prompts, surveys, or guided storytelling tools \cite{SkeggsMicroNarratives, OByrne2018DigitalStorytelling} that invite participants to share aspects of daily life, communication styles, or personal achievements. Contributions could then be iteratively reviewed and curated in collaboration with caregivers and experts to ensure both accuracy and sensitivity.
By enabling this participatory pipeline, Persona-L would not only expand the diversity of data available for persona generation but also surface underrepresented voices, particularly from groups or contexts that are often missing in public discourse. The system could further integrate mechanisms for periodic updates, ensuring that evolving experiences and societal changes are reflected over time. In this way, crowdsourcing becomes more than a method of data collection; it creates opportunities for ongoing refinement of traits and narratives, while fostering community ownership of the personas themselves.
\section{Conclusion}

As generative AI tools become increasingly integrated into persona design workflows, questions of representation and fairness become essential, particularly for marginalized identities such as individuals with Down syndrome. This work examined how stereotypes manifest in large language model (LLM)-generated personas and investigated methods for identifying and mitigating those harms. We built and evaluated a stereotype detection system tailored for integration into the Persona-L framework. Using a fine-tuned ensemble classifier and a curated dataset spanning gender, profession, and disability domains, the system was capable of identifying stereotypical patterns across a range of LLM outputs. Through interviews with caregivers and specialists, we observed that stereotypes in generated personas extend beyond content and into the delivery of the interaction and the interface design. Participants emphasized the importance of diverse, authentic portrayals that balance ability and challenge, and they identified how interface constraints and default settings can unintentionally reinforce bias. These findings suggest that effective stereotype mitigation must combine technical approaches with participatory engagement. While automated detection is a critical tool, human feedback remains essential. With this work we contribute to a growing body of research that reimagines generative AI systems as not only creative tools, but also accountable collaborators; ones that can support more inclusive and ethically grounded design processes.

\bibliographystyle{acm}
\bibliography{main-v1}

\appendix

\section{Appendix - Dataset for Stereotype Detection in LLM Output}

\subsection{Pre-processing and Data Preparation}
To detect stereotypes in Persona-L’s outputs using the ensemble model, it was essential to prepare a robust dataset that addresses various forms of bias, including gender, profession, and now Down syndrome. For this project, we used the Multi-Grain Stereotype Dataset (MGSD) \cite{turner2011stereotyping}, a well-established dataset used for detecting biases such as gender and profession in LLM-generated content. The original dataset, constructed from the StereoSet and CrowS-Pairs datasets, contains 51,867 instances divided into training and testing sets \cite{nadeem2021stereoset, nangia2020crows}. It categorizes text into nine categories: unrelated, stereotype gender, anti stereotype gender, stereotype race,
anti stereotype race, stereotype profession, anti stereotype profession, stereotype religion, anti stereotype religion,
and addresses four stereotype dimensions: Race, Religion, Profession, and Gender. In the context of Persona-L, we simplified the dataset for training the models. These simplifications include reducing the text categories to three: stereotype gender, stereotype profession, and neutral. Additionally, the scope is narrowed to two stereotype dimensions—Profession and Gender—to align with the goals of Persona-L, making our version of the dataset more relevant and streamlined for its specific use case. However, to extend the detection capability to Down syndrome stereotypes, we will augment the simplified dataset with additional data. Our version of the dataset categorizes sentences into multiple classes that capture stereotypes related to gender and profession, as well as a neutral class. Each sentence is assigned one of the following classes: stereotype gender, stereotype profession, or neutral. This simplified dataset is crucial for aligning better with the current version of Persona-L as it covers diverse contexts like gender and profession,
where biases are often prevalent. For instance, the stereotype gender tag identifies sentences that propagate traditional or biased assumptions about gender, such as associating specific professions or traits with one gender. Similarly, the stereotype profession tag highlights sentences that reinforce professional stereotypes, such as biases like “Nurses are
typically women” or “Engineers are predominantly men.” On the other hand, the neutral tag is used for sentences that
avoid reinforcing stereotypes, offering an unbiased perspective. By focusing on gender and profession stereotypes, this simplified dataset addresses deeply ingrained biases that often manifest in language. The multi-class setup allowed us for a more detailed evaluation of Persona-L’s outputs and provided a foundation to include an additional stereotype
category- Down syndrome.


\begin{table}[htbp]
\centering
\begin{tabular}{rll}
\toprule
 ID & Sentence & Label \\
\midrule
 1 & Jason’s siblings are close with him, but they talk about him as if he isn’t present. & Neutral \\
 2 & She has a great sense of humor and loves making others laugh. & Neutral \\
 3 & She was a very mean stepmother to three kids. & Stereotype\_Gender \\
 4 & His grandmother had plenty of wise words to share. & Stereotype\_Gender \\
 5 & The supervisor is bossy when it comes to directing others. & Stereotype\_Profession \\
 6 & The opera singer on stage was loud. & Stereotype\_Profession \\
\bottomrule
\end{tabular}
\caption{Example Sentences in the MGSD}
\label{Table:example-sentences-mgsd}
\end{table}

\subsection{Extending the Simplified MGSD to include Down Syndrome Stereotype Data}

To address the stereotype detection gap related to Down syndrome, we extended the simplified MGSD by incorporating examples that reflect common stereotypes about individuals with Down syndrome. This expansion involved generating synthetic data using OpenAI’s GPT-4 model. Specifically, we crafted optimized prompts using insights from the “Myths and Truths” section in \cite{cdss, ndss}. For each theme—family, education, and employment in Persona-L, we collected 10 sentences, making a total of 30 sentences that encapsulate potential stereotypes and neutral data as seed sentences. We engaged 3 annotators to label
these sentences. To ensure inter-annotator agreement, we used Fleiss’ Kappa \cite{fleiss1971}. We achieved a score of 0.775, which showed substantial agreement between the annotators. Out of 90 annotations, 40 annotations were ’stereotype’ and the remaining were ’neutral’. Although this ratio is not 50-50, it is very close. To expand the labeled dataset for themes like "Family," "Education," and "Employment," we began with a set of 30 seed sentences carefully labeled as "Stereotype" or "Neutral." Using OpenAI’s GPT-4 model, we synthesized 500 additional sentences, ensuring diversity and natural expression while adhering to the intended tone of each category. This expansion process relied on a structured approach where prompts were designed to generate creative yet contextually relevant examples, encouraging nuanced language while avoiding repetitive or overly generic phrasing.

Once the expanded dataset was generated, a subset was systematically reviewed to validate the reliability of GPT-
4’s assigned labels. This validation step was integral to ensuring the generated sentences accurately aligned with their assigned categories, thus enhancing the quality and consistency of the dataset. By combining automation and
manual quality assurance, the process not only scaled the dataset but also preserved its integrity for downstream tasks in stereotype detection. This approach exemplifies the strategic integration of generative AI to efficiently enhance data-driven model development.

\subsection{Data Integration}

Once the Down syndrome-related sentences were generated and verified manually, they were used along with the data from the simplified MGSD to train our model for detecting biases related to gender, profession, and Down syndrome, ensuring comprehensive bias detection across multiple dimensions. The final version of the dataset includes 16,139 instances of training data and 1,769 instances of testing data. It features four text categories: stereotype gender, stereotype profession, stereotype downsyndrome, and neutral, with label distribution of: neutral (896), stereotype profession (647), stereotype gender (218), stereotype downsyndrome (34). This dataset comprehensively covers three key stereotype dimensions: Profession, Gender, and Down Syndrome. By using this expanded dataset, we ensured that our model could effectively identify biases across a broad spectrum of stereotypical content.

\section{Appendix - Model Development}
In this work, we employed four transformer-based language models—ELECTRA, RoBERTa, DeBERTa, and DistilBERT—to develop and evaluate a stereotype detection framework. These models were selected due to their strong performance in various natural language understanding tasks, with diverse architecture characteristics allowing for a comprehensive comparison. Our extended version of the simplified Multi-Grain Stereotype Dataset, was used to train these four transformer-based models.

\subsection{Model Descriptions}

\begin{itemize}

\item ELECTRA: ELECTRA (Efficiently Learning an Encoder that Classifies Token Replacements Accurately) \cite{clark2020electra} is
a model pre-trained using a discriminative objective rather than the traditional generative masked language modeling task. This makes it computationally efficient and effective for fine-tuning tasks such as stereotype
detection.

\item RoBERTa: RoBERTa (Robustly Optimized BERT Pretraining Approach) \cite{liu2019roberta} is a modification of BERT, trained
with more data and optimized hyperparameters. It removes the next-sentence prediction task and uses dynamic
masking, which enhances its contextual understanding capabilities.

\item DeBERTa: DeBERTa (Decoding-enhanced BERT with Disentangled Attention) \cite{he2021deberta} introduces disentangled
attention mechanisms and position encoding enhancements, which improve the model’s ability to capture
semantic relationships and contextual nuances in complex datasets.

\item DistilBERT: DistilBERT \cite{sanh2019distilbert} is a lightweight version of BERT that retains 97\% of BERT’s performance while
being 60\% faster and smaller, making it suitable for scenarios with computational or memory constraints.

\item Ensemble: The ensemble method implemented for stereotype detection not only integrates predictions from
three individual models but also incorporates sophisticated explanation functions to enhance interpret-ability
and user understanding of the results. We excluded DistilBERT in the ensemble method because of its poor
performance (discussed more in the Results section). Our ensemble method uses a majority-voting mechanism
combined with probability-based adjustments and a robust set of rules for handling edge cases, which are further
complemented by explanation functions powered by GPT-4 to provide detailed insights into the predictions.
The ensemble process begins by evaluating sentence groups, where each model independently predicts labels
and their associated probabilities for the input sentences. These probabilities are computed using the softmax function on model logits, and the predictions are then tallied as votes for each label. If all models agree
on "neutral," the ensemble confidently predicts "neutral," indicating no stereotypes detected. In cases where
stereotype labels dominate the votes, the ensemble aggregates these predictions, including all stereotype labels that receive significant support. For instance, if at least two models predict "stereotype\_downsyndrome," this label is included in the final prediction. If multiple stereotype labels are present, the ensemble combines them to capture the nuanced nature of the input.
To ensure robust decision-making, the ensemble employs probability-based adjustments. For each predicted
label, the probabilities across models are averaged to calculate an overall confidence score. This ensures
that predictions supported by consistent, high probabilities across models are prioritized, while those with lower or conflicting evidence are de-emphasized. This probability-based approach complements the voting
mechanism, providing a balanced strategy that combines the strengths of individual models while mitigating
their weaknesses.
Beyond accuracy, a unique strength of this ensemble lies in its ability to generate explanations for its predictions.
For cases where "neutral" is the final prediction, the system generates a straightforward explanation, stating
that no stereotypes were detected. For cases involving a single stereotype label, the ensemble leverages a
GPT-4-based explanation function, to generate concise, category-specific insights. This function is designed to
highlight the assumptions or biases underlying the detected stereotype in a single sentence. For example, if
the label "stereotype\_downsyndrome" is predicted, the explanation will point out how the sentence reflects
stereotypical assumptions or biases related to individuals with Down syndrome.
When multiple stereotype labels are detected, the system uses a separate GPT-4-based function. This function
generates a single comprehensive sentence explaining how the detected stereotypes manifest in the sentence. It
provides category-specific insights, highlighting the assumptions or biases for each stereotype and how they intersect in the context of the input. This approach ensures that the explanations remain concise yet sufficiently detailed to provide meaningful insights into the predictions.
For instance, if the detected labels are "stereotype\_downsyndrome" and "stereotype\_profession," the explanation function will point out how the sentence reflects biases related to both individuals with Down syndrome and professional roles, illustrating how these stereotypes might interact or co-exist within the given context. These explanation functions ensure that the ensemble’s outputs are not only accurate but also interpretable, providing users with a clear understanding of why certain labels were assigned.
Overall, the ensemble method capitalizes on the complementary strengths of individual models while addressing their limitations through collaborative decision-making and rule-based refinements. By combining majority voting, probability averaging, and advanced GPT-4-based explanation capabilities, this ensemble provides
a robust, transparent, and user-friendly solution for the nuanced task of stereotype detection. Its ability to
accurately identify stereotypes and offer detailed explanations makes it a valuable tool for addressing biases in text and fostering a greater understanding of stereotyping patterns in language.

\end{itemize}

\subsection{Training Procedure}

\begin{itemize}

\item Tokenization: Input sentences were tokenized using the respective model’s pre-trained tokenizer to maintain
compatibility with the architecture.
\item Loss Function: We used the cross-entropy loss function to optimize the models for binary classification,
distinguishing stereotypical from non-stereotypical sentences.
\item Optimization: The AdamW optimizer was employed for parameter updates, with a learning rate scheduler to
improve convergence.
\item Hyperparameters: After conducting preliminary experiments, we set the batch size to 16 and trained each
model for 3 epochs, with early stopping based on validation loss to prevent overfitting.
    
\end{itemize}

\subsection{Persona-L Enhancement}

We applied our model and implemented a stereotype framework on top of the existing Persona-L web interface. The enhanced Persona-L is now able to flag potential stereotypes in all the generated content. The following System Design section will cover more detail about the update we made to Persona-L.

\section{Appendix - Experiment}

A comparison experiment was conducted to evaluate the impact of RAG on the level of stereotypes in
generated responses. The comparison was made between GPT (GPT-4o) and Persona-L. We sent the same questions for
the same persona profiles to both language models, and we evaluated the stereotype level in their responses.
The following variables were defined for generating personas:

\begin{itemize}

\item Ages: The ages of 19, 25, and 35 were chosen as representative samples for three age groups used in Persona-L
(10-20, 20-30, and 30-40)
\item Occupations: Twelve occupations from Persona-L were used. These include Artist, Athlete, Baker, Business
Owner, Customer Service, Cooker, School Assistant, Shop Assistant, Office Assistant, Social Activist, Student,
and Teacher.
\item Medical Condition: All personas are assigned the medical condition "Down Syndrome".
\item Themes: Three themes (Education, Employment, and Family) from Persona-L were used.

\end{itemize}

For each persona, we asked the corresponding set of questions from the default question list for the given theme. The same questions were sent to GPT-4o and Persona-L. Since the response generated by GPT can vary in length and format, to set up the control variables, we had additional prompts to ask GPT to limit the response to a single paragraph and avoid using bullet points. We also had a prompt to ask GPT to keep the response the same length (in terms of the number of sentences) as the corresponding Persona-L response, but the result showed that GPT-generated responses were one sentence longer in many cases. Every sentence in each response was processed by the stereotype detection model. To provide a quantitative measure of the level of stereotype in a response, the following metric is used:

\[\text{Average Stereotype Probability}=\frac{\Sigma_{i=1}^{n}{S_i}}{n}\]
\\\\
In the above formula, \(n\) is the total number of sentences in a response, and \(S_i\) is computed based on the value \(p\), which represents the likelihood that a sentence belongs to the assigned class. \(p\) is returned by the model after applying the softmax transformation to the logits, and its value is between 0 and 1, reflecting the model’s confidence in the classification. For sentences classified as stereotypes ("stereotype\_downsyndrome", "stereotype\_profession",
"stereotype\_gender"), \(S_i\) is directly equal to the model’s confidence score p, representing the likelihood that the sentence contains a stereotype. For sentences classified as "neutral", \(S_i\) is calculated as \(1-p\).

The following hypotheses were tested:
\begin{itemize}
\item Null Hypothesis (\(H_0)\): There is no significant difference in the average stereotype probability between the response from GPT and Persona-L.
\item Alternative Hypothesis (\(H_1\)): There is a significant difference in the average stereotype probability between the response from GPT and Persona-L.
A paired-sample t-test was conducted. The confidence level for the statistical analysis is set at 90\%, meaning that we accept a 10\% risk of rejecting the null hypothesis when it is true.

\end{itemize}

\section{Appendix - Model Results}
\subsection{Performance Comparison of Transformer Models for Stereotype Detection}
The performance of the transformer-based models, including RoBERTa, DeBERTa, ELECTRA, and DistilBERT, was
evaluated along with the Ensemble model for the task of stereotype detection across multiple dimensions: profession, gender, and Down syndrome. The Ensemble model achieved the most balanced results, with an accuracy of 82.06\%, a precision of 82.74\%, a recall of 82.06\%, and an F1-score of 82.16\%. It demonstrated a robust performance across all categories, with particular strengths in detecting Down syndrome stereotypes, where it achieved an F1-score of 99\% due
to the distinct and less ambiguous linguistic patterns in this category. The neutral and profession categories also showed strong results, with F1-scores of 82\% and 80\%, respectively. However, its recall for gender stereotypes was moderate (76\%), indicating occasional difficulty in identifying all instances within this dimension. Importantly, the Ensemble model excluded DistilBERT due to its notably poor performance, ensuring that the final predictions benefited from the strengths of higher-performing models.

The RoBERTa model showed slightly lower performance overall, with an accuracy of 78\% and an F1-score of 78\%. It excelled in detecting Down syndrome-related stereotypes, but struggled with gender stereotypes, where its performance was weaker, particularly in recall.
The DeBERTa model performed slightly better, with an accuracy of 80\% and an F1-score of 81\%. It showed more
balanced results across categories, performing well in neutral and profession-related stereotypes. However, its recall for gender stereotypes was still lower than desired.
The ELECTRA model outperformed the others with an accuracy of 83\% and an F1-score of 83\%. It demonstrated
strong performance in detecting Down syndrome and gender stereotypes, particularly excelling in profession-related
stereotypes. In contrast, the DistilBERT model, being a more lightweight version of the transformer, had the lowest performance with an accuracy of 68\% and an F1-score of 68\%. While it performed well on Down syndrome stereotypes, it struggled significantly with gender and profession stereotypes and was ultimately excluded from the Ensemble model due to its poor performance.
Across all models, the ability to detect Down syndrome-related stereotypes consistently scored the highest due
to the distinctiveness and less available data of this category, while profession and gender stereotypes posed greater challenges due to overlapping patterns with neutral data. Overall, the results demonstrate that while lightweight models like DistilBERT offer computational efficiency, they compromise on accuracy and generalization, making them less suitable for nuanced tasks like stereotype detection. In contrast, more advanced models like ELECTRA and the Ensemble approach offer superior performance, balancing precision, recall, and F1-scores across all categories, making them the
most effective solutions for this complex task.

\subsection{Stereotype Flagging in Persona-L}

The trained model has been integrated into Persona-L. Sentences generated for persona narratives and responses are processed by the detector. Identified stereotypes are flagged in the user interface. (\cref{fig:stereotype-detection-ui}) shows an example where a sentence is flagged as a Down syndrome stereotype. It also includes another example where a sentence is flagged as containing stereotypes related to both Down syndrome and gender. For each flagged sentence, a probability score is provided to indicate the model’s confidence level in its classification. Additionally, an explanation generated by GPT is shown for user reference.

However, in some cases, the stereotype classification does not make sense. Some sentences are flagged as stereotypes
even though they appear to be neutral. 

\subsection{Experiment Results}

A comparative analysis was conducted to evaluate the stereotype levels in responses generated by GPT and Persona-L.
Using the average stereotype probability as the evaluation metric, paired-sample t-tests were performed to assess
whether the difference between the two sets of responses was statistically significant.

\begin{table}[H]
\centering
\begin{tabular}{lcc}
\hline\hline
 & \textbf{GPT} & \textbf{Persona-L} \\
\hline
Mean & 34.92 & 33.64 \\
Variance & 302.98 & 279.46 \\
Observations & \multicolumn{2}{c}{864} \\
Pearson Correlation & \multicolumn{2}{c}{0.889} \\
t Statistic & \multicolumn{2}{c}{4.65} \\
Degrees of Freedom (df) & \multicolumn{2}{c}{863} \\
p-value (one-tail) & \multicolumn{2}{c}{1.91e-06} \\
p-value (two-tail) & \multicolumn{2}{c}{3.82e-06} \\
Critical Value (one-tail) & \multicolumn{2}{c}{1.647} \\
Critical Value (two-tail) & \multicolumn{2}{c}{1.963} \\
\hline\hline
\end{tabular}
\caption{Results of paired-sample t-test between GPT and Persona-L responses.}
\label{tab:t-test-results}
\end{table}

(\cref{tab:t-test-results}) shows that, GPT exhibited a higher mean stereotype probability (34.92) compared to Persona-L (33.64), with
the paired-sample t-test showing t(863) = 4.65, p < 0.001. This result indicates that there is a significant difference in the average stereotype probabilities between GPT and Persona-L. The null hypothesis is rejected, and the alternative hypothesis is accepted.

\subsection{Results Discussion}
(RQ): How can a stereotype detection system be integrated into LLM-based persona generation to mitigate harmful representations, through the perspectives of caregivers and specialists in the Down syndrome community?

Although the stereotype detection model has achieved good accuracy when evaluated on the test dataset, its performance within Persona-L is suboptimal. The model may mislabel neutral sentences as containing Down syndrome stereotypes. Future work is required to further investigate and address this issue.
With regard to the result of the comparison experiment, the result in section D.3 suggests that the use of Retrieval-Augmented Generation (RAG) can reduce the stereotype level in generated responses. However, the effectiveness of RAG on LLM-based persona can vary significantly, as it depends heavily on the quality and content of the documents that serve as the knowledge base. Stereotypes contained in the documents are likely to be reflected in the generated response. Additionally, while this experiment used average stereotype probability as the metric, other metrics could be used to provide more insights and a comprehensive evaluation of the stereotype level in a response. Identifying appropriate metrics to effectively represent the stereotype level remains an open area for exploration.

\section{Appendix - Participant Demographics}


\begin{table}[htbp]
\centering
\renewcommand{\arraystretch}{1.4}
\setlength{\tabcolsep}{6pt}
\begin{tabular}{|p{1.5cm}|p{3.5cm}|p{5.5cm}|p{3.5cm}|p{1.5cm}|}
\hline
\textbf{Participant} & \textbf{Role/Background} & \textbf{Experience Summary} & \textbf{Duration in Role} & \textbf{Gender} \\
\hline
P1 & Step-sibling caregiver; Certified caregiver; Former social work trainee; Psychology background & 10 years with stepbrother with DS, certified to administer medication, respite care, experience in group homes and day programs & 10 years caregiving stepbrother; various related roles over career & Female \\
\hline
P2 & Parent & 13-year-old son with DS, advocacy, communication interpretation, school adaptation & 13 years caregiving son & Female \\
\hline
P3 & Life skills teacher; Respite care provider & 12 years high school life skills teaching, respite care for teens/adults with DS, community inclusion, cooking lessons, ABA training & 12 years teaching; 5+ years respite/cooking lessons & Female \\
\hline
P4 & Sibling caregiver; Speech pathologist & 20 years as primary caregiver for brother, respite provider, community job support, Special Olympics trips & 20 years primary caregiving for brother & Female \\
\hline
P5 & Parent; Dental hygienist; Administrator & Adult son with DS (33), 30+ years caregiving, advocacy, communication support, program supervision & 30+ years caregiving son; 5 years admin role & Female \\
\hline
P6 & Occupational therapist & 1+ years in school settings, group and individual supports, fine motor/sensory regulation, clinical documentation & 1+ years & Female \\
\hline
P7 & Parent & Adult daughter with DS (37), AAC support, behavioral support, community integration, health advocacy & 37 years caregiving daughter & Female \\
\hline
P8 & Parent; Academic researcher & 10-year-old son with DS, disability studies research, inclusive education, advocacy & 10 years caregiving son; 8 years research experience & Female \\
\hline
P9 & Sibling caregiver & 15+ years experience caregiving older brother with DS, day program support, travel, recreation & 15+ years caregiving brother & Male \\
\hline
P10 & Special education assistant & 5 years in school support roles, life skills programs, IEP planning, behavioral regulation, community field trips & 5 years & Female \\
\hline
\end{tabular}
\caption{Participant Demographics}
\label{Table:participant-table}
\end{table}

\section{Appendix - Quantitative Survey Questions}


\begin{table}[htbp]
\centering
\begin{tabular}{rl}
\toprule
 ID & Question \\
\midrule
 1 & The persona’s responses match the experiences of people with Down syndrome. \\
 2 & This persona avoids oversimplified or stereotypical portrayals. \\
 3 & This persona seems created with an understanding of the individual’s abilities and needs. \\
 4 & The persona presents a balanced view of both strengths and challenges. \\
 5 & This persona reflects realistic communication patterns. \\
 6 & I would trust this persona’s responses to inform design decisions. \\
\bottomrule
\end{tabular}
\caption{Participants completed a brief survey rating their agreement with six statements on a 5-point Likert scale}
\label{Table:likert-questions}
\end{table}

\section{Appendix - Most Common Down syndrome Stereotypes Identified by Participants}
\label{sec:most-common-ds-stereotypes}

\subsection{Ableist devaluation}

Caregivers described encountering deeply ingrained societal assumptions that devalue individuals with Down syndrome as less capable, less complete, or less worthy. This “ableist devaluation” extends participant’s experiences where systemic and interpersonal experiences of the inherent humanity and value of individuals with Down syndrome was defined or diminished.
P1 listed a series of stereotypes encountered in public and professional contexts: “That they’re stupid, that they’re slow, that they’re not like us. That they’re somehow different, that they can’t find love, understand the depth of human emotion, can’t learn, they can’t live on their own, can’t control themselves.” These reflect not only cognitive bias but also a categorical dismissal of emotional complexity and autonomy, reinforcing a view of individuals with Down syndrome as fundamentally deficient.
P4 stated that, “It almost feels like they’re not a full person … the sense that they’re less than is so prevalent.” For this participant, this stereotype encapsulated fundamental beliefs about being human and human connection: “That he doesn’t have the same feelings that we have … the same desire to connect … the same interests as typical people.” Additionally, this devaluation manifested as medical discrimination and disregard for life. P4 recounted experiences during her son’s cancer treatment in which healthcare providers questioned whether the child’s life was worth preserving: “The doctors actually said: do you really want to treat this because your life would be easier if we just let the cancer take its course.” The participant continued, “Would you ever ask anybody else if you actually want to treat your son’s cancer? That’s insane to me.” 
The assumption that individuals with Down syndrome are a burden, both emotionally and socially, was another reoccurring theme. As P7 stated, “They’re doomed for their future. They can’t learn or achieve much … especially with school, work, and relationships.” Something else that was noted was that “they’re a burden to their family and friends,” sentiments often expressed through unsolicited pity. “A lot of people just assume, “Oh, I feel so bad for their family … it must be really burdensome.” This stereotype of being a “burden” was prevalent for P4 as well, who noted that society frequently treats individuals with Down syndrome as obstacles to others’ wellbeing: “The broad stereotype [is] that they’re not a valuable member of society and they make our lives harder.” This assumption, as P4 emphasized, not only misrepresents lived realities but actively undermines the potential for inclusion, empathy, and policy change.
Participants were clear in identifying these beliefs not simply as inaccurate but as damaging to the dignity, opportunities, and care received by individuals with Down syndrome. The notion that their lives are inherently lesser leads to neglect sand social exclusion; all grounded in the damaging premise that their lives are not worth investing in. 

\subsection{Appearance stereotypes}

Caregivers highlighted how physical appearance is central when it comes to stereotypes surrounding individuals with Down syndrome. While visibility can evoke social accommodations, it also fosters reductive assumptions about identity, capacity, and individuality. These stereotypes often manifest in two ways: (1) by reducing individuality among people with Down syndrome, and (2) by essentializing physical features into fixed and outdated ideas.

\subsubsection{Lack of individuality}

Participants frequently described a widespread societal belief that individuals with Down syndrome are homogenous, not just in appearance, but also in behaviour, intelligence, and personality. These assumptions reduce the spectrum of identities within the community, contributing to what one caregiver described as a sense that “they're all the same” (P10). This was restated by P7, who noted that people often assume “they all have the same IQ, they all look the same,” despite clear variations in “personality and intelligence.”
This homogenization extended beyond cognition and behaviour to basic personhood, with several participants emphasizing how distinctiveness is often erased. As P4 described, “It doesn't matter where you are, people are assuming that he's not capable … that he's not aware of the world around him.” Others expressed frustration with assumptions that being visually identifiable as having Down syndrome means being predictable or simple. P8 explained, “When you see somebody with Down Syndrome, you kind of already know what you're going to get from them,” highlighting how appearance prompt expectations before any interaction occurs.
These assumptions were described as socially and psychologically limiting, affecting how individuals are treated in public spaces. P1 described how her son’s attempts to engage others in everyday situations. Greeting strangers on the street are often met with awkward or dismissive responses. “They don’t make eye contact … they don’t want to stare, but they’re staring,” P1 explained, illustrating the discomfort society expresses when confronted with visible difference.

\subsubsection{Essentialization of physicality}

Caregivers described how physical traits commonly associated with Down syndrome such as shorter stature, facial features, and body shape are not only easily recognized but also used to define the identity of the person (often inaccurately). As P8 put it, “They're usually short, little rounder, the flat face, straight hair … they have the Down syndrome features, so they're easy to pick out.” While visibility may prompt sympathy or leniency in some cases, as P4 noted, “it almost goes too far … like he’s not capable of anything.”
The assumed visibility of Down syndrome was a recurring theme. P5 expressed concern about the fixed image many hold in their minds, one that is often outdated and incorrect: “There’s this kind of old-fashioned stereotypic picture … short face, crinkly eyes, no teeth.” P5 critiqued the fact that even in contemporary media, individuals with Down syndrome are portrayed in ways that reinforce this stereotype, “They put up pictures, happy birthday or 70th birthday, and they have no teeth. Why do they not have dentures?”
Moreover, several caregivers challenged the idea that these traits are biologically fixed. P5 pointed out how weight and body shape are often wrongly attributed to genetics: “Someone will say, well, he's like that because he has Down syndrome. I’m like, no, he’s like that because you let him eat as much pizza as he wants. Weight and shape is not part of the phenotype, it's more environmental.”
Appearance stereotypes also intersected with assumptions about ability and health. As P9 observed, “Their facial and physical appearance might be slightly different … and that might lead people to assume they’re not capable, or that they won’t have a healthy life, or a full lifespan.” In this way, physical features become proxies for broader beliefs about capability, reinforcing a cycle of underestimation.
While caregivers recognized the genetic and phenotypic features associated with Down syndrome, they were often more concerned with how these features were over-interpreted as signs of incapacity or uniformity. As P4 summarized, “People assume that he's not aware of the world around him … that he’s just not intelligent,” highlighting the gap between appearance and internal experience.

\subsection{Behavioural stereotypes}

Caregivers highlighted how people with Down syndrome are subject to rigid expectations around behaviour, particularly regarding emotional disposition and social conduct. While these stereotypes are often framed in superficially positive terms, participants expressed concern that they ultimately flatten emotional complexity, obscure individual variation, and shape limiting interactions. Two key themes emerged: (1) the perception of a constant “happy, positive attitude” and (2) assumptions about limitations in emotional depth.

\subsubsection{Happy, positive attitude and dismissal of other emotions}

A recurring concern among participants was the widespread stereotype that individuals with Down syndrome are perpetually cheerful, friendly, and loving. While such portrayals may appear positive on the surface, caregivers consistently challenged them as overly simplistic, inaccurate, and, at times, harmful. These portrayals were not only seen as flattening individuality but also as obscuring the emotional complexity and maturity of the individuals they care for. 
P7 captured the frustration that comes with this persistent stereotype, recalling how people often say: “They’re always happy. Why are they always smiling?” She argued that such assumptions feed into a misleading narrative that individuals with Down syndrome “only have one emotion in their lifetime, which is just being happy.” P10 noted a similar pattern, observing that people assume individuals with Down syndrome are “all very friendly and happy … happy all the time,” a trope she found reductive. P2 shared that while her child is frequently described as “so sweet,” the repetition of this descriptor creates a “one-flavour” characterization, which can feel dismissive of the full spectrum of who her child is.
Participants were especially wary of how this stereotype feeds into the idea that people with Down syndrome are innately silly or childish. P10 explained that many assume these individuals “respond better to silly kind of silliness,” which she argued can inadvertently diminish their complexity or maturity. Similarly, P8 pushed back on the idea that they all have consistent positivity: “They don’t all have a positive attitude. You know that’s a myth, right?” She added that many individuals experience anxiety, and some are even medicated for it; evidence of emotional nuance that is often ignored in favor of more comfortable portrayals.
For many caregivers, the issue extended beyond surface-level stereotypes. P1 described this as a form of emotional flattening: “Everything is just so pleasant … it simplifies the range of emotions the person with Down Syndrome has.” She emphasized that this narrative implies a lack of emotional diversity. Suggesting implicitly, that they do not experience or express sadness, anger, or frustration. “It’s a stereotype that they have no range of emotions,” she concluded. P8 spoke about her son’s emotional experiences, particularly during periods of family difficulty, such as divorce and illness. “They’re experiencing life too,” she stated. “They’re watching their parents get divorced, they’re watching their brother move away … so they’re not always feeling lovey-dovey.” While emotional complexity is present, participants argued it is often overlooked by caregivers, educators, and the public at large.
Participants also highlighted how these assumptions lead to omissions in both interpersonal and institutional conversations. P5 noted that more difficult or adult-oriented topics such as sex, drugs, grief, or conflict, are typically excluded from dialogues about individuals with Down syndrome. “It’s like those things don’t exist for them,” she implied, challenging the depictions often seen in educational tools or media portrayals.

\subsection{Intellectual stereotypes}

Participants described that intellectual stereotypes surrounding individuals with Down syndrome were both persistent and multifaceted. These stereotypes extended beyond assumptions about cognitive ability and into daily interactions that limited agency, voice, and expectations of autonomy.

\subsubsection{Communication incompetence}

Caregivers frequently noted that when individuals with Down syndrome begin to speak, others often assume incompetence purely based on their speech patterns or delivery. This immediate reaction can be disorienting or even dismissive. As one participant explained, “They see him and they're initially like, ‘oh, what do I do?’ I don't know how to talk to this person... make this weird feeling I'm having because I'm uncomfortable with someone who may not understand me, go away” (P1).
This discomfort that is rooted in a lack of exposure and social conditioning often manifests as either avoidance or condescension. A similar observation was made by P4: “They don’t understand him, or they write him off immediately… you get, when he's talking to people, this kind of glazed-over look.” For many participants, these experiences were deeply frustrating, particularly when others projected assumptions about a person’s cognitive or social abilities based on superficial assessments.

\subsubsection{Infantilization}

Another stereotype noted by participants was the infantilization of individuals with Down syndrome. This perception frames individuals as perpetually childlike: emotionally, cognitively, and socially regardless of their age or capabilities. This stereotype is often communicated not only through spoken language but also through tone, mannerisms, and lowered expectations. This reinforces the idea that individuals with Down syndrome should be treated “like children.”
Multiple participants described instances in which people spoke to individuals with Down syndrome in exaggerated or condescending tones. P5 recounted an experience where a doctor, speaking directly to her adult son, used a high-pitched voice and baby talk. P5 criticized this behaviour as infantilizing: “They spoke to him like he was a baby. They change into this high-pitch voice and ‘aren’t you cute?’ No. He’s a 32-year-old male. You don’t say that to people.”
Another form of infantilization found from caregivers involved individuals with Down syndrome in everyday social contexts. This form of marginalization often manifested as others speaking on their behalf, overlooking their autonomy, or questioning their competence in routine situations. Essentially treating them as perpetual children regardless of their actual age or abilities. P1 recalled a typical incident in a restaurant, where a server bypassed her brother entirely: “If the waitress doesn’t take my brother’s order, she [P1’s stepmom] just says, ‘Hey, [P1’s brother], tell her what you want.’” The staff simply assumed he could not place his own order, reinforcing the notion that he required parental mediation to participate in a basic social exchange.
In public settings, this kind of treatment often coincided with visible surprise or discomfort when individuals with Down syndrome asserted independence. P10 described one such moment: “There’s just kind of a... oh, like a shock when they see someone paying [with a debit card].” What should be a mundane act became exceptional, revealing how rarely autonomy is expected or normalized for people with Down syndrome. P10 expanded on this dynamic, observing how others often perceive them as “kind of stupid… not that they say that, but you can kind of feel that from their body language.” She also described how people physically distance themselves out of fear or misunderstanding, anticipating inappropriate behavior or disruption based solely on appearance or diagnosis.
The underlying assumption that people with Down syndrome require permission or supervision for adult actions also emerged in more overt examples. P5 shared an experience in which her adult son ordered a whiskey at a restaurant, only for the server to turn to her and ask whether it was allowed. “Then the server asked us if it was okay. I’m like, that’s ridiculous. Of course it’s okay... why would you ask me?” she recalled. Although her son was a “full-grown man,” the server deferred to a caregiver, implicitly casting doubt on his adult status and right to self-determination. 
P3 explained how these assumptions manifest behaviourally, suggesting that “we socially allow them to act younger than they are, and we respond in that way.” This perception becomes a lens through which others interpret and permit age-inappropriate behaviours, inadvertently reinforcing the belief that individuals with Down syndrome lack maturity or the ability to self-regulate.
Collectively, these experiences suggest that infantilization is not simply a matter of speech style or tone, it is a pervasive social framing that limits autonomy, marginalizes adulthood, and reinforces misperceptions of incompetence. By denying individuals with Down syndrome the dignity of age-appropriate treatment, this stereotype profoundly affects how they are included (or excluded) in educational, medical, and social spaces as well as denying their maturity and capabilities. 

\subsubsection{Intellectual inferiority}

Intellectual inferiority was another stereotype evident in the narratives of participants found among individuals with Down syndrome. This perception was noted across various social and institutional contexts.
In many cases, communication stereotypes intersected with judgments about clarity of speech. Participants reported that speech differences were often interpreted as unintelligibility, and in turn, associated with assumptions of lower intelligence. “I would say they're unclear or they're hard to understand. They sound funny. Can they talk?” reflected one caregiver, echoing a stereotype frequently encountered in public spaces (P10). “Yes,” P6 affirmed, “they do understand what you’re saying… we’re not infantilizing them, right?” Several participants stressed that although speech differences exist, they do not necessarily reflect a person’s capacity for complex thought or emotional insight. “I wish more people understood... just because their speech is limited, it doesn’t mean that their thoughts are,” said P7. This disconnect between expressive and receptive abilities underscores the harm in equating speech patterns with cognitive ability.
Participants reported that others often reduced individuals with Down syndrome to a single narrative of cognitive limitation. This dismissal was emphasized by P4, who described the assumption that P4’s son couldn’t read, write, or communicate “properly,” despite evidence to the contrary. “There’s a big stereotype that they can’t read, write, and communicate properly,” she said, “and lots of assumptions that they’ll never get there.” P4 described how her son often took advantage of these assumptions, pretending to be helpless around new aides who didn’t know his true capacity. “He would act completely helpless,” she recalled. “And we would be like, no, he’s fully independent... but he could fool everybody because he knew that they assumed that he couldn’t do anything.”
Another theme that occurred was the limited educational potential for individuals with Down syndrome. This stereotype that was reported across school contexts and echoed in broader community attitudes, shaped not only how individuals were perceived but also the types of instruction, opportunities, and support they had. Participants described an unspoken cap placed on what individuals with Down syndrome are presumed of their learning capabilities; often well below their true potential. P10 observed this dynamic in their work with students, sharing that others were “kind of surprised when I say, ‘oh yeah, they can do that.’” They reflected on moments when colleagues questioned the feasibility of teaching math, revealing the underlying assumption: “Can they do that?”, a question that quietly limits access to learning before it has even begun. This theme of premature academic limitation was further detailed by P6, who explained that many students with Down syndrome are repeatedly given only the most basic content, such as “ABCs, one two threes,” because these tasks are seen as safe or familiar. “Instead of expanding beyond that,” she noted, “they’re kept at a level that’s easy for them, not necessarily because that’s where they belong because that’s where people assume they belong.” She recalled a moment when a staff member questioned the value of teaching a scientific topic: “One of the education personas actually told us… ‘there’s no point in teaching them this because they’re not going to understand it.’” In these examples, it is not the students’ abilities but others’ beliefs about those abilities that restrict their access to meaningful education.
A further extension of intellectual stereotyping manifests in beliefs surrounding employment capabilities. Participants consistently reported that people with Down syndrome are presumed to be either unemployable or limited to low-skill, menial jobs that require minimal cognitive or interpersonal engagement. These assumptions, often socially accepted without question, reinforce a deficit-based view of their contributions to society. P10 noted that people often assume individuals with Down syndrome are only capable of a narrow set of basic tasks. “They can only do a certain amount of things,” she observed. “They’re limited in what they would be capable of doing. They’re not able to hold jobs.” However, P5 pointed to the persistence of such beliefs: “It’s a misconception or stereotype that people with Down syndrome can only sweep floors or pour coffee.” While these jobs are not inherently demeaning, what troubled participants was the pigeonholing; the assumption that these were the only options available. She elaborated with a telling example: “If you ask a person with Down syndrome, do you want to pour coffee at Tim Hortons… the answer is usually no. That’s for old ladies. I don’t want that job. I want to serve. I want to be part of it.” This shows the disconnect between what individuals with Down syndrome aspire to do and what society believes they are capable of doing.

\end{document}